\documentclass{aastex701}

\usepackage{xcolor}

\newcommand{\edit}[1]{#1}

\newcommand{\aot}{$A\!\cdot\!\Omega$}   
\newcommand{\degsq}{deg$^2$}
\newcommand{\photoz}{photo-$z$}
\newcommand{\Msun}{M_\odot}

\shorttitle{Rubin 2036 Workshop}
\shortauthors{Plazas Malagón et al.}

\graphicspath{{./}{figures/}}

\begin{document}

\title{The \emph{Rubin 2036: Opportunities for Discovery} Workshop:
A Community Roadmap for the Future of the
Vera C.\ Rubin Observatory After the LSST}


\newcommand{\SLAC}{SLAC National Accelerator Laboratory}
\newcommand{\KIPAC}{Kavli Institute for Particle Astrophysics and Cosmology, Stanford University}
\newcommand{\Rubin}{NSF--DOE Vera C. Rubin Observatory}
\newcommand{\NOIRLab}{NSF NOIRLab}
\newcommand{\Stanford}{Stanford University}
\newcommand{\UChicago}{University of Chicago}
\newcommand{\Berkeley}{University of California, Berkeley}
\newcommand{\LBNL}{Lawrence Berkeley National Laboratory}
\newcommand{\Kent}{University of Kent}
\newcommand{\IPN}{ESFM, Instituto Polit\'ecnico Nacional}
\newcommand{\Fermilab}{Fermi National Accelerator Laboratory}
\newcommand{\Princeton}{Princeton University}
\newcommand{\Pittsburgh}{University of Pittsburgh}

\author[orcid=0000-0002-2598-0514]{Andr\'es Alejandro Plazas Malag\'on}
\altaffiliation{Editor and corresponding author}
\affiliation{\SLAC}
\affiliation{\KIPAC}
\affiliation{\Rubin}
\email[show]{plazas@stanford.edu}

\author[orcid=0000-0003-0667-5941]{Steve Allen}
\affiliation{\KIPAC}
\affiliation{\SLAC}
\email{swa@stanford.edu}

\author[orcid=0000-0003-3768-7515]{Shreya Anand}
\affiliation{\Stanford}
\email{sanand08@stanford.edu}

\author[orcid=0000-0003-0930-5815]{Timo Anguita}
\affiliation{Universidad Andr\'es Bello, Santiago, Chile}
\email{timo.anguita@unab.cl}

\author[orcid=0009-0000-9174-8492]{Muhammad Awais}
\affiliation{\IPN}
\email{mawais2500@alumno.ipn.mx}

\author[orcid=0009-0009-6932-6379]{Henry Best}
\affiliation{Masaryk University, Brno, Czech Republic}
\email{hbest@mail.muni.cz}

\author[orcid=0000-0003-3195-5507]{Simon Birrer}
\affiliation{Stony Brook University}
\email{simon.birrer@stonybrook.edu}

\author[orcid=0000-0001-7665-5079]{Lindsey Bleem}
\affiliation{Argonne National Laboratory}
\email{lbleem@anl.gov}

\author[orcid=0000-0002-8622-4237]{Robert Blum}
\affiliation{\NOIRLab}
\affiliation{\Rubin}
\email{bob.blum@noirlab.edu}

\author[orcid=0000-0002-9836-603X]{Adam S. Bolton}
\affiliation{\SLAC}
\email{abolton@slac.stanford.edu}

\author[orcid=0000-0001-9297-7748]{Rosaria Bonito}
\affiliation{INAF, Osservatorio Astronomico di Palermo, Italy}
\email{rosaria.bonito@inaf.it}

\author[orcid=0000-0001-6966-5316]{Alex Broughton}
\affiliation{\SLAC}
\affiliation{\Stanford}
\affiliation{\Rubin}
\email{abrought@stanford.edu}

\author[orcid=0000-0002-3936-9628]{Jeffrey L. Carlin}
\affiliation{\Rubin}
\affiliation{\NOIRLab}
\email{jeff.carlin@noirlab.edu}

\author[orcid=0000-0001-7316-4573]{Francisco Castander}
\affiliation{Institute of Space Sciences (ICE-CSIC), Barcelona}
\email{fjcastander@gmail.com}

\author[orcid=0000-0002-3925-7802]{Eric Charles}
\affiliation{\KIPAC}
\affiliation{\SLAC}
\email{echarles@slac.stanford.edu}

\author[orcid=0000-0003-3917-0966]{Rebecca C. Chen}
\affiliation{\KIPAC}
\affiliation{\SLAC}
\email{rcchen0@stanford.edu}

\author[orcid=0000-0002-7155-679X]{Anirudh Chiti}
\affiliation{\Stanford}
\email{achiti@stanford.edu}

\author[orcid=0000-0001-5386-0133]{Lyla Choi}
\affiliation{\KIPAC}
\email{lc3535@princeton.edu}

\author{Chuck Claver}
\affiliation{\NOIRLab}
\affiliation{\Rubin}
\email{cclaver@lsst.org}

\author[orcid=0009-0006-2305-7183]{Suber Corley}
\affiliation{Arizona State University}
\email{sdcorle1@asu.edu}

\author[orcid=0000-0002-2495-3514]{John Franklin Crenshaw}
\affiliation{\KIPAC}
\affiliation{\SLAC}
\email{jfcren@stanford.edu}

\author[orcid=0000-0002-4928-4003]{Arjun Dey}
\affiliation{\NOIRLab}
\email{arjun.dey@noirlab.edu}

\author[orcid=0000-0002-5296-4720]{Seth Digel}
\affiliation{\KIPAC}
\affiliation{\SLAC}
\email{digel@stanford.edu}

\author[orcid=0000-0002-7790-9971]{Holger Drass}
\affiliation{\Rubin}
\email{hdrass@lsst.org}

\author[orcid=0000-0001-8251-933X]{Alex Drlica-Wagner}
\affiliation{\Fermilab}
\affiliation{\UChicago}
\email{kadrlica@fnal.gov}

\author[orcid=0000-0002-1080-0955]{Haruki Ebina}
\affiliation{\Berkeley}
\email{ebina@berkeley.edu}

\author[orcid=0000-0003-3918-7995]{Lukas Eisert}
\affiliation{\SLAC}
\email{eisert@slac.stanford.edu}

\author[orcid=0000-0003-2314-5336]{Anthony Englert}
\affiliation{Brown University}
\email{anthoyn_englert@brown.edu}

\author[orcid=0000-0001-5717-2688]{Sydney Erickson}
\affiliation{University of California, Los Angeles}
\email{serickson@astro.ucla.edu}

\author[orcid=0000-0003-2371-3356]{Kevin Fanning}
\affiliation{\SLAC}
\email{fanning@slac.stanford.edu}

\author[orcid=0000-0003-4079-3263]{Josh Frieman}
\affiliation{\Fermilab}
\affiliation{\UChicago}
\email{jfrieman@uchicago.edu}

\author[orcid=0000-0001-5422-1958]{Shenming Fu}
\affiliation{\SLAC}
\affiliation{\KIPAC}
\affiliation{\Rubin}
\email{sfu@slac.stanford.edu}

\author[orcid=0000-0002-3870-1537]{Nandini Hazra}
\affiliation{National Centre for Nuclear Research (NCBJ), Warsaw}
\email{nandini.hazra@ncbj.gov.pl}

\author[orcid=0000-0002-4043-6445]{Carrie E. Holt}
\affiliation{Las Cumbres Observatory}
\email{cholt@lco.global}

\author[orcid=0000-0003-4738-4251]{Minhee Hyun}
\affiliation{\KIPAC}
\affiliation{\SLAC}
\affiliation{\Rubin}
\affiliation{Korea Astronomy and Space Science Institute}
\email{minhee51@stanford.edu}

\author[orcid=0009-0005-5796-6054]{Hafiz Inam Ullah}
\affiliation{\IPN}
\email{inambaloch383@gmail.com}

\author[orcid=0000-0001-8738-6011]{Saurabh Jha}
\affiliation{Rutgers University}
\email{saurabh@physics.rutgers.edu}

\author[orcid=0000-0002-6825-5283]{J. Bryce Kalmbach}
\affiliation{\SLAC}
\affiliation{\KIPAC}
\email{brycek@slac.stanford.edu}

\author[orcid=0000-0002-5261-5803]{Yijung Kang}
\affiliation{\SLAC}
\email{ykang@slac.stanford.edu}

\author[orcid=0000-0002-8130-3593]{Kshitija Kelkar}
\affiliation{Indian Institute of Technology Bombay}
\email{kkshitija.astro@gmail.com}

\author[orcid=0000-0001-9821-4987]{Sthabile Kolwa}
\affiliation{\KIPAC}
\affiliation{University of South Africa}
\email{sthabile.kolwa@gmail.com}

\author[orcid=0000-0001-5139-1978]{Andjelka Kova\v{c}evi\'c}
\affiliation{University of Belgrade}
\email{andjelka.kovacevic@matf.bg.ac.rs}

\author[orcid=0000-0001-8967-2281]{Kenneth W. Lin}
\affiliation{\LBNL}
\affiliation{\Berkeley}
\email{kwlin@berkeley.edu}

\author[orcid=0000-0003-2908-4597]{Meng-Xiang Lin}
\affiliation{Simon Fraser University}
\email{mengxiang_lin@sfu.ca}

\author{Becky Long}
\affiliation{\Kent}
\email{rl454@kent.ac.uk}

\author[orcid=0000-0002-9514-7245]{Sean MacBride}
\affiliation{Physik-Institut, University of Zurich, Winterthurerstrasse 190, 8057 Zurich, Switzerland}
\email{sean.macbride@physik.uzh.ch}

\author[orcid=0000-0002-8676-1622]{Alex I. Malz}
\affiliation{Space Telescope Science Institute}
\email{aimalz@nyu.edu}

\author[orcid=0000-0001-8205-9441]{Steve Margheim}
\affiliation{\NOIRLab}
\email{steven.margheim@noirlab.edu}

\author[orcid=0000-0002-0113-5770]{Philip J. Marshall}
\affiliation{\SLAC}
\affiliation{\KIPAC}
\email{pjm@slac.stanford.edu}

\author[orcid=0000-0002-4468-8324]{Nigel J. Mason}
\affiliation{\Kent}
\email{n.j.mason@kent.ac.uk}

\author[orcid=0000-0001-6013-1131]{Guillem Megias Homar}
\affiliation{California Institute of Technology}
\email{gmegias@caltech.edu}

\author[orcid=0000-0002-2849-559X]{Simona Mei}
\affiliation{Universit\'e Paris Cit\'e}
\affiliation{CNRS-IN2P3}
\email{mei@apc.in2p3.fr}

\author[orcid=0000-0001-7964-9766]{Hironao Miyatake}
\affiliation{Kobayashi-Maskawa Institute (KMI), Nagoya University}
\email{hironao.miyatake@nagoya-u.jp}

\author{Maia Moore}
\affiliation{\Kent}
\email{TBD}

\author[orcid=0000-0003-4446-7465]{Ver\'onica Motta}
\affiliation{Universidad de Valpara\'iso, Chile}
\email{veronica.motta@uv.cl}

\author[orcid=0000-0001-6145-5859]{Justin Myles}
\affiliation{\Princeton}
\email{jmyles@astro.princeton.edu}

\author[orcid=0000-0001-8684-2222]{Jeffrey A. Newman}
\affiliation{\Pittsburgh}
\email{janewman@pitt.edu}

\author[orcid=0009-0008-0111-1809]{Anh Nguyen}
\affiliation{\Stanford}
\email{qanh@stanford.edu}

\author[orcid=0000-0001-7966-7606]{Yasna Ordenes-Brice\~no}
\affiliation{Universidad Diego Portales}
\email{yasna.ordenesbriceno@gmail.com}

\author[orcid=0000-0002-6011-0530]{Antonella Palmese}
\affiliation{Carnegie Mellon University}
\email{palmese@cmu.edu}

\author[orcid=0000-0002-5854-7426]{Swayamtrupta Panda}
\affiliation{International Gemini Observatory}
\affiliation{\NOIRLab}
\email{swayamtrupta.panda@noirlab.edu}

\author[orcid=0000-0003-0512-5489]{Claire Poppett}
\affiliation{\Berkeley}
\affiliation{\LBNL}
\email{clpoppett@lbl.gov}

\author[orcid=0000-0002-6716-4179]{Francisco Pozo Nu\~nez}
\affiliation{Heidelberg Institute for Theoretical Studies (HITS)}
\email{Francisco.Pozonunez@h-its.org}

\author[orcid=0000-0001-5999-7923]{Anand Raichoor}
\affiliation{APC, CNRS}
\affiliation{\LBNL}
\email{raichoor@apc.in2p3.fr}

\author[orcid=0009-0000-3218-9846]{Andrew Rasmussen}
\affiliation{\SLAC}
\email{arasmus@slac.stanford.edu}

\author[orcid=0009-0005-8281-1644]{Maya Redden}
\affiliation{\KIPAC}
\email{msredden@stanford.edu}

\author[orcid=0000-0002-0418-6258]{Bernardita Ried Guachalla}
\affiliation{\KIPAC}
\affiliation{\SLAC}
\email{bried@stanford.edu}

\author[orcid=0000-0003-1301-9221]{Steve Ritz}
\affiliation{University of California, Santa Cruz}
\email{sritz@ucsc.edu}

\author[orcid=0000-0001-5326-3486]{Aaron Roodman}
\affiliation{\SLAC}
\affiliation{\KIPAC}
\email{roodman@slac.stanford.edu}

\author[orcid=0000-0002-5307-5941]{M\'onica Rubio}
\affiliation{Universidad de Chile}
\email{mrubio@das.uchile.cl}

\author[orcid=0000-0002-8502-8947]{Leonid Sajkov}
\affiliation{\KIPAC}
\email{sajkov@stanford.edu}

\author[orcid=0000-0002-6909-205X]{Paras Sharma}
\affiliation{Stony Brook University}
\email{paras.sharma@stonybrook.edu}

\author[orcid=0009-0006-6662-5056]{Ben Sherwin}
\affiliation{\Stanford}
\affiliation{\SLAC}
\email{sherwinb@stanford.edu}

\author[orcid=0000-0003-0973-4900]{Brian Stalder}
\affiliation{\NOIRLab}
\affiliation{\Rubin}
\email{brian.stalder@noirlab.edu}

\author[orcid=0000-0003-0347-1724]{Chris Stubbs}
\affiliation{Harvard University}
\email{stubbs@physics.harvard.edu}

\author[orcid=0000-0002-0896-8134]{John Su\'arez-P\'erez}
\affiliation{Tecnol\'ogico de Monterrey}
\email{jf.suarez@tec.mx}

\author[orcid=0000-0002-9121-3436]{Sandrine Thomas}
\affiliation{\NOIRLab}
\affiliation{\Rubin}
\email{sandrine.thomas@noirlab.edu}

\author[orcid=0000-0001-7836-2261]{Chun-Hao To}
\affiliation{\UChicago}
\email{chto@uchicago.edu}

\author[orcid=0009-0001-9603-2618]{Alessandro S. Tramuto}
\affiliation{Universit\`a degli Studi di Palermo (UNIPA), Italy}
\affiliation{INAF, Osservatorio Astronomico di Palermo, Italy}
\email{alessandrosalvatore.tramuto@community.unipa.it}

\author[orcid=0000-0001-6161-8988]{Yousuke Utsumi}
\affiliation{National Astronomical Observatory of Japan}
\email{yousuke.utsumi@nao.ac.jp}

\author[orcid=0000-0002-4115-0318]{Laura Venuti}
\affiliation{SETI Institute}
\email{lvenuti@seti.org}

\author[orcid=0000-0001-8459-1036]{Hai-Feng Wang}
\affiliation{LuoxiaHong Institute of Astronomy}
\email{haifeng.wang.astro@gmail.com}

\author[orcid=0000-0003-2229-011X]{Risa Wechsler}
\affiliation{\KIPAC}
\affiliation{\SLAC}
\email{rwechsler@stanford.edu}

\author[orcid=0000-0003-2919-7495]{Christina C. Williams}
\affiliation{\NOIRLab}
\email{christina.williams@noirlab.edu}

\author[orcid=0000-0001-7113-1233]{W. Michael Wood-Vasey}
\affiliation{\Pittsburgh}
\email{wmwv@pitt.edu}

\author[orcid=0009-0003-2770-5537]{Prince Yadav}
\affiliation{University of Illinois Urbana-Champaign}
\email{princey2@illinois.edu}

\author[orcid=0000-0003-1585-997X]{Masaya Yamamoto}
\affiliation{\Princeton}
\email{masayamamotou@gmail.com}

\begin{abstract}
The NSF-DOE Vera C.\ Rubin Observatory will complete its ten-year Legacy Survey of Space and Time (LSST) in the mid-2030s. Its 8.4\,m aperture and 9.6\,\degsq\ field of view give it an \'etendue unmatched by any facility planned for that era. The \emph{Rubin 2036: Opportunities for Discovery} workshop (KIPAC/SLAC, 3--5 August 2026) brought together $\sim$100 participants to ask \edit{what should be done with Rubin after the LSST}. This paper summarizes the workshop's structure and discussions. Five working groups considered continued LSST-style surveying; alternative cadences and footprints; new filters; targets of opportunity and multi-messenger astronomy; and enabling technologies and upgrades. The continued-surveying group explored \edit{a machine-learning ``foundation model,'' pretrained on Rubin's own imaging,} for detection, deblending, and classification in the faint, crowded sky. The cadence group described a process for allocating post-LSST observing time, along with cadences for a transiting-exoplanet census, Type~Ia supernova cosmology, time-delay cosmography, and second-scale transients. The filter group explored offset broadband, medium-band, and narrowband filters for photometric redshifts, high-redshift large-scale structure, AGN reverberation mapping, Galactic archaeology, and emission-line imaging. The multi-messenger group explored target-of-opportunity follow-up of next-generation gravitational-wave and neutrino sources, including standard sirens, kilonovae, and Galactic supernovae. The technology group organized instrumental options by cost and lead time, from configuration changes such as fast readout modes to moderate upgrades such as new filters, a faster shutter, and summit edge computing. Cross-cutting topics included AI/ML, Chilean and Latin-American co-leadership, satellite megaconstellations, and weather and long-term climate trends at Cerro Pach\'on. The working groups' white papers are available through the workshop's Zenodo community (\url{https://zenodo.org/communities/rubin2036workshop/}) and the workshop website, which also includes workshop talks (\url{https://rubin-observatory.github.io/Rubin-Beyond-LSST/pages/Rubin2036proceedings/}).
\end{abstract}


\section{Introduction}
\label{sec:intro}

The Vera C.\ Rubin Observatory was designed and built to carry out the Legacy
Survey of Space and Time (LSST): a decade-long, \edit{wide, fast, deep (WFD)} optical survey
of the southern sky in six broad photometric bands ($ugrizy$) \citep{ivezic2019, bianco2022}. The four
science pillars of the LSST are probing dark energy and dark matter, taking an
inventory of the Solar System, exploring the transient \edit{and variable} sky, and mapping
the formation and structure of the Milky Way. The LSST began in 2026, and by the
mid-2030s the nominal ten-year survey will be complete, delivering an
unprecedented \edit{deep, six-band ($ugrizy$), time-domain data set spanning} $\sim$18{,}000\,\degsq. The telescope,
camera, data-management system, and the software will remain one \edit{of} the most capable wide-field imaging \edit{systems} worldwide.

This paper
summarizes the outcomes of the \emph{Rubin 2036: Opportunities for Discovery}
workshop\footnote{\url{https://rubin-observatory.github.io/Rubin-Beyond-LSST/}}
held at the Kavli Institute for Particle Astrophysics and Cosmology
(KIPAC)\footnote{A joint institute of SLAC National Accelerator Laboratory and
Stanford University.} on 3--5 August 2026. It presents \edit{potential directions} for Rubin
operations beyond the LSST, built on concrete science cases that the community could develop for upcoming prioritization
processes, including the next decadal survey in astronomy and astrophysics
(Astro2030).

The scientific cases for Rubin rely on its unique \'etendue (\aot, the product of
collecting area and field of view). No facility planned for the 2030s will match
Rubin's combination of an 8.4\,m aperture and a 9.6\,\degsq\ field of view\edit{, together with rapid telescope slewing and short settling times that make it efficient for both wide-area and rapid target-of-opportunity observing}. Other
imaging and spectroscopic facilities, such as the Dark Energy Spectroscopic
Instrument (DESI), the 4-meter Multi-Object Spectroscopic Telescope (4MOST), the
\emph{James Webb Space Telescope} (JWST), the \emph{Nancy Grace Roman Space
Telescope} (Roman), and \emph{Euclid}, will provide complementary capabilities. By
the mid-2030s we will also have \edit{a large suite} of data products and scientific
discoveries enabled by the LSST itself.

Continuing to operate and evolve a built facility \edit{can be} a high-return use of
resources: the marginal cost of new capability is small relative both to the cost
already committed and to any new facility of comparable reach.

Over the past decade a series of community studies has argued for
planning Rubin's use beyond the LSST, from the Cosmic Visions Dark Energy reports
\citep{cvde_science, cvde_technology, cvde_smallprojects}, through Astro2020 white
papers on next-generation LSST science and future uses of the facility
\citep{jha2019_nextgen, kahn2019_futureuses}, the LSST next-generation
instrumentation workshop \citep{nextgen_instrumentation2019}, dedicated
narrow-band filter studies \citep{yoachim2019_narrowband}, and the Snowmass 2021
process \citep{blum2022_rubinafterlsst, snowmass_complementarity, snowmass_cf04} (see Appendix~\ref{app:prior} for a summary of previous work). The Astro2020 decadal survey, \emph{Pathways to Discovery in Astronomy and Astrophysics for the 2020s} \citep{astro2020} set the current
scientific priorities for the field\edit{. Its} recommendations are now the subject of a
National Academies mid-decadal review of progress \citep{astro2020_midreview}, and
\edit{considering science drivers now for the post-LSST future of Rubin} positions the community to bring
well-developed science cases to the next decadal survey
(Astro2030).

The workshop brought together approximately 100 in-person and online participants
to begin developing this roadmap. A program of invited framing talks on the first
day and synergy talks on the second (Appendix~\ref{app:talks}), together with short
\edit{`lightning'} talks (Appendix~\ref{app:lightning}), introduced the
scientific, programmatic, and cross-facility context and seeded many of the science
cases; slides for a subset are available on the workshop website.
The bulk of the time was devoted to collaborative work in five thematic groups: (A) continued
LSST-style surveying (Section~\ref{sec:groupA}); (B) alternative cadences and
footprints (Section~\ref{sec:groupB}); (C) narrow- and medium-band surveys and
alternative filters (Section~\ref{sec:groupC}); (D) targets of opportunity and
multi-messenger astronomy (Section~\ref{sec:groupD}); and (E) enabling technology
and upgrades (Section~\ref{sec:groupE}). Each group developed and compared science
cases against a common set of prompts, reproduced in full in
Appendix~\ref{app:prompts}\edit{:} the specific science goal
and the single result that would \edit{matter most}; why Rubin, and why \edit{the decision to pursue the capability} cannot
wait, framed around its area $\times$ field-of-view (\aot); the
strongest skeptic's objection and a rebuttal; the concrete change required (filter
set, cadence, instrument, or software) and its lead time; which early-LSST data
release must be in hand \edit{before the case is compelling}; the success metric and figure of merit \edit{for the case}; connections to
other facilities, surveys, and groups; and the single most useful next step. Two cross-cutting perspectives, the role of artificial intelligence and machine learning (AI/ML) and opportunities for Latin-American, particularly Chilean, co-leadership (Section~\ref{sec:chile}), were considered across all cases. Other cross-cutting topics, such as satellite megaconstellations and climate-related degradation of seeing, emerged organically during the meeting. The themes were informed by discussions within the Scientific Organizing Committee, participant input during registration, and the previous community efforts noted above.

Each thematic group developed its results into a standalone document, archived
in the workshop's Zenodo community; these are listed in Table~\ref{tab:zenodo} and
cited at the start of each group's section below, and the present paper is a
summary that points to them for full detail.

\begin{table*}[t]
\centering
\caption{The five per-group documents produced by the workshop and archived in
the Rubin~2036 Zenodo community. Full titles, author lists, and DOIs are given in
the cited references.}
\label{tab:zenodo}
\begin{tabular}{c l l}
\hline
Group & Title (Zenodo community) & Reference \\
\hline
A & \parbox[t]{3.9in}{\raggedright A survey-native foundation model for the faint, crowded sky} & \citet{yadav2026_deepdisc} \\[4pt]
B & \parbox[t]{3.9in}{\raggedright Cosmology and Astrophysics with the Vera C. Rubin Observatory Using Alternative Cadences and Footprints} & \citet{rubin2036_groupB} \\[4pt]
C & \parbox[t]{3.9in}{\raggedright Cosmology and Astrophysics with the Vera C. Rubin Observatory Using New Filters} & \citet{rubin2036_groupC} \\[4pt]
D & \parbox[t]{3.9in}{\raggedright Multi-Messenger Cosmology and Astrophysics after the LSST} & \citet{rubin2036_groupD} \\[4pt]
E & \parbox[t]{3.9in}{\raggedright Enabling Technology and Upgrades for Rubin 2036} & \citet{rubin2036_groupE} \\
\hline
\end{tabular}
\end{table*}

The workshop program and supporting material are collected on the workshop website, {\url{https://rubin-observatory.github.io/Rubin-Beyond-LSST/}} which is intended to serve as a persistent home for this and future meetings in the series. The documents from each working group are available in the Rubin~2036 Zenodo community (\url{https://zenodo.org/communities/rubin2036workshop}).

\section{Group A: Continued LSST-style Surveying}
\label{sec:groupA}

\noindent\textit{Contributors:} P.~Yadav. See \citet{yadav2026_deepdisc}
for the full Group~A document.\\

Group~A considered continuing to survey
with the \emph{current} system beyond the nominal ten years, gaining scientific
reach from a longer time baseline, and deeper coadds rather than from any new
hardware \edit{or change of observing strategy}. Its case argues that the deeper, more crowded imaging a continued
survey produces can only be exploited if the detection, deblending, and
classification software evolves in step, an argument for sustained investment in
\edit{data-analysis software, computing, and machine-learning infrastructure (including graphics-processing-unit (GPU) compute and curated training data)} rather than in the \edit{observatory} itself.

\subsection{A survey-native foundation model for the faint, crowded sky}
\label{sec:A-ml}

A continued, deepening survey \edit{produces science only} if its faint, crowded sky
can actually be measured, and at those depths the fraction of blended detections
rises steeply while the availability of vetted training labels does not. The
central question is whether a \emph{survey-native foundation model} (a backbone
pretrained self-supervised on Rubin's own unlabeled $ugrizy$ coadds) can deliver
scene-level detection, deblending, and star/galaxy classification at the faint
end using only a small fraction of the labels that supervised training requires.
The single result that would matter is matching fully supervised performance
from roughly 5--15\% of the labels, making a scene-level catalog deployable
across the whole footprint rather than only in the handful of
spectroscopically rich deep fields. The \aot\ argument here is two-sided: the
same survey depth that drives fields into the blended regime (where
morphology-only star/galaxy separation degrades to $\sim$50\% correctly
classified injected stars at $i=23.8$ in \edit{Rubin} Data Preview 1 (DP1) source-injection
tests) also guarantees that unlabeled pixels, unlike vetted labels, scale
exactly with \aot. Evidence from the first application of the DeepDISC \edit{\citep{merz2023_deepdisc}}
scene-level framework to real LSST DP1 imaging (1{,}458 six-band tiles over the
Extended Chandra Deep Field South and the Euclid Deep Field South; 288{,}277 objects, validated against the Hubble Space Telescope
and the Cosmic Assembly Near-infrared Deep Extragalactic Legacy Survey) shows scene-level inference reaching 94.8\%/71.4\% galaxy/star
recall against HST (near-parity with a photometry-based random forest) while
achieving 94.6\% catalog purity on a held-out HST footprint versus 77.6\% for
the Scarlet-based \edit{\citep{melchior2018_scarlet}} production pipeline on the same sky. The requirement is not
telescope time but infrastructure: sustained GPU
compute and bulk-image access for data-release-scale pretraining
($\sim$2--3\,yr to stand up), a deep-field truth program (source-injection plus
curated spectroscopic/astrometric labels, coordinated with DESI/4MOST), and an
integration path to release scene-level catalogs as supported value-added data
products. The priors, rerunning the $i=23.8$ injection benchmark with
image-based classifiers and \edit{a sweep over the fraction of labeled training examples}, are obtainable from
Data Preview 2 (DP2) and \edit{Data Release 1 (DR1)}. This case \edit{is cross-cutting},
connecting to \emph{Roman} (joint pixel-level training cuts the simulated
unrecognized-blend fraction from 18.6\% to 11.6\%), \emph{Euclid}, DESI and \edit{Spec-S5 (a proposed Stage-V spectroscopic facility)}, and to
Groups~B (which deep fields exist to train on), C (photo-z quality is bounded by
blended-photometry quality), D (host identification in crowded fields), and E
(compute/software infrastructure). A Chilean co-leadership tie-in was proposed
around the truth/label program and a data-science training-school module.

\section{Group B: Alternative Cadences and Footprints}
\label{sec:groupB}

\noindent\textit{Contributors:} H.~Best, S.~Birrer,
R.~Blum, A.~S.~Bolton, R.~Bonito,
R.~C.~Chen, S.~Corley, S.~Digel,
S.~Erickson, C.~E.~Holt, S.~Jha, A.~I.~Malz,
P.~Marshall, P.~Sharma, A.~S.~Tramuto, L.~Venuti, W.~M.~Wood-Vasey,
P.~Yadav; and Chilean co-leads T.~Anguita and V.~Motta. See \citet{rubin2036_groupB} for the full Group~B document.\\

Group~B addressed how to re-tune Rubin's observing strategy after LSST (e.g., denser
cadence, \edit{micro-,} mini- and sub-surveys, monitoring programs, and target-of-opportunity
splits) to reach science that the homogeneous, compromise-driven \edit{WFD} main-survey
cadence cannot. Group B participants also identified a ``metacase'' for the workshop at large: a proposal that the
community establish a flexible, prioritized \emph{process} for allocating
post-LSST observing time, rather than commit now to a single fixed alternative
cadence. Around that umbrella the group developed a set of concrete
science-driven cadences.

\subsection{The metacase: a process for the post-LSST science portfolio}
\label{sec:B-meta}

The observation motivating the metacase is that Rubin's capabilities are not
exhausted by the LSST survey design: survey-cadence proposals have consistently
oversubscribed the available LSST time, and new ideas \edit{will keep emerging} as the
survey progresses. Because cadence and footprint changes have \edit{shorter lead
times than any other facility modification} (shorter than filters or focal-plane hardware),
they \edit{would not involve downtime for observatory updates and could in fact be pursued on an interim basis if, for example, new filters were being fabricated}. What Rubin needs is a survey-strategy framework
that can run concurrent programs of differing scale, area, filter set, exposure
time, and cadence, together with a scientific prioritization process, modeled
on the large/medium/small proposal structure used at other observatories (e.g.,
\emph{Roman}'s mix of core community surveys and proposal-based general-astrophysics
surveys). The counterargument that the LSST cadence was already optimized
in a community process is answered by noting that the most important science
in 2036 will differ from the original LSST science goals, and that the main
survey necessarily compromised across all cases rather than serving finely
targeted ones. Success would be measured by the number and oversubscription of
proposals received and by the breadth of community participation; a Chilean
tie-in (reserved Chilean-community time or a Chilean Time Allocation Committee,
TAC) is a natural component.
The metacase is the umbrella under which most Group~B, and several cross-group,
cases sit. \edit{One concrete instantiation discussed was a dedicated time-domain observing mode that relaxes the main survey's uniformity and large-area requirements in order to maximize Rubin's
time-domain legacy.}

\subsection{Science-driven cadences}
\label{sec:B-cases}

The group developed several cases that such a process would prioritize:

\paragraph{Transiting-exoplanet census of the Milky Way disk}

A census of transiting planets across all spectral types in the Galactic disk,
confirming candidates through $\geq3$ transits, over $\sim$3{,}600\,\degsq\ of
the disk. It requires operational flexibility rather than hardware: \edit{a high-cadence mode that revisits each field many times per night on 1--6\,hr cycles}.
Kepler (0.25\% of sky) and the Transiting Exoplanet Survey Satellite (TESS;
bright stars, $r\lesssim12$) leave most of the disk's planet population
unprobed; a benchmark target is confirming planets around $\geq$10{,}000 disk
stars.

\paragraph{Young stellar objects across all timescales}
A multi-band ($ugriz$), high-cadence (30-minute), short-duration ($\sim$10-day)
sub-survey of thousands of young stellar objects (YSOs) in several star-forming
regions of Rubin's Galactic footprint would link short-term accretion bursts to
long-term eruptive events such as those of EXor-type stars (young stars that
undergo repeated months-to-years-long accretion outbursts). Exploiting Rubin's \'etendue and its blue
and red filters, it would recover light-curve morphologies, determine stellar
photospheric properties, and characterize accretion physics as a function of
stellar age and environment. Its complementarity with the low-cadence, multi-year
LSST view uniquely constrains the occurrence rates of small- versus large-scale
events in the accretion histories of young stars, extending the YSO micro-survey
already proposed for the LSST \citep{bonito2023_ysolsst}.

\paragraph{Type Ia supernova cosmology to $z\sim1.1$}
A dedicated sub-survey drilling into non-Deep-Drilling-Field regions to fill the
Hubble diagram out to $z\sim1.1$, where the ten-year LSST sample is sparse,
\edit{which would enable} tests of time-varying dark energy beyond $(w_0,w_a)$ and tests of
cosmic isotropy along continuous sightlines. It needs no new filters; Group~C's
offset filters would most benefit this case by preserving discovery depth.
Mid-to-late-release cosmology results are the prior that would set the strategy.

\paragraph{Strongly-lensed AGN and supernovae for time-delay cosmography}
Daily cadence on $\sim$2{,}000\,\degsq, preferentially
on the \emph{Roman} high-latitude footprint, to measure time delays of lensed quasars
and the rare, rapidly-fading lensed supernovae. Concrete performance metrics
include accretion-disk sizes measured to 10\%, a $5\sigma$ discrimination \edit{between competing models of the
lens-population stellar initial mass function (IMF)}, and 1\% $H_0$ from each of
the lensed-SN and lensed active-galactic-nucleus (AGN) samples.

\paragraph{The few-second sky: ultracompact binaries and fast transients}
Two cases exploit second-scale cadence. A fast-cadence Galactic survey
(1--2\,s continuous imaging over an hour, repeated over years) targets
Laser Interferometer Space Antenna (LISA)-era eclipsing double white dwarfs
with sub-15-minute periods, comparing the optically-timed orbital decay $\dot
P$ against the gravitational-wave (GW)-driven prediction to isolate tidal
dissipation and test whether such binaries are calibrated GW sirens, a
measurement whose precision scales as $\sim T^{-5/2}$, so \edit{a baseline beginning in the mid-2030s would maximize sensitivity by the time of LISA's late-2030s Galactic-binary
catalog}. A complementary extragalactic case targets fast blue optical
transients and fast-radio-burst (FRB) counterparts, which are washed out in a
standard 30\,s visit and require 1--2\,s readouts plus second-scale alerting.
Both depend on
enabling technology from Group~E (fast readout modes).

\paragraph{Solar-System and coordinated-observing mini-surveys}
An ecliptic-restricted, daily-cadence mini-survey pairing $g$ and $r$ within each
visit would catch the rise of cometary and active-asteroid outbursts and
distinguish dust- from gas-driven activity by color; and a framework for
commensal or simultaneous observing with other facilities (the Dark Energy
Camera, DECam; the Southern Astrophysical Research Telescope, SOAR; the Very
Large Telescope, VLT; LS4; radio: Simons Observatory, the Atacama Cosmology
Telescope, ACT; the Square Kilometre Array, SKA) would enable science requiring
same-time multi-wavelength coverage (explicitly excluding multi-messenger
triggers, which are Group~D's domain).

\section{Group C: Narrow- and Medium-band Surveys and Alternative Filters}
\label{sec:groupC}

\noindent\textit{Contributors:} M.~Awais, L.~Bleem,
A.~Broughton, J.~Carlin, F.~Castander,
E.~Charles, A.~Chiti, J.~F.~Crenshaw,
A.~Dey, A.~Drlica-Wagner, H.~Ebina,
A.~Englert, K.~Fanning, M.~Hyun,
J.~B.~Kalmbach, K.~Kelkar, S.~Kolwa,
A.~Kova\v{c}evi\'c, R.~Long, A.~I.~Malz,
S.~Margheim, N.~J.~Mason, S.~Mei,
H.~Miyatake, M.~Moore, J.~Myles,
J.~Newman, A.~Nguyen,
Y.~Ordenes-Brice\~no, S.~Panda,
F.~Pozo Nu\~nez, A.~Raichoor, A.~Rasmussen,
M.~Redden, L.~Sajkov, B.~Sherwin,
S.~Thomas, C.-H.~To, H.~I.~Ullah,
C.~Williams, M.~Yamamoto. \\

Group~C examined multiple ways of changing
Rubin's filter complement beyond the Sloan-heritage $ugrizy$ set (which is
optimized for \photoz\ near $z\sim0.2$): medium-band filters
($\lambda/\Delta\lambda\sim15$--20, $\sim$200--800\,\AA\ wide) for photometric
redshifts, high-$z$ cosmology, AGN reverberation
mapping, and stellar archaeology; narrow-band filters for emission-line and
biosignature-proxy surveys; ``offset'' broad bands shifted in wavelength
relative to $ugrizy$ while preserving depth and cadence; and more exotic hardware
concepts (ramp filters, a segmented ``butcher-block'' tiled filter array, and
exploitation of the bandpass--angle-of-incidence coupling of Rubin's telecentric
design). \edit{The group's document organizes its cases by science theme, from
cosmology and large-scale structure to Local-Universe and Galactic archaeology,
exoplanets and astrobiology, and Solar System science; we follow that grouping
below.} A
recurring observation was that essentially every proposal costs of order one to a
few million dollars per filter, a small marginal cost on an existing facility.
The unifying \edit{metric} proposed was the ratio
$R(T,\edit{\Omega})=\mathrm{FoM}(\text{new filter set})/\mathrm{FoM}(ugrizy)$ at fixed time
$T$ and \edit{sky area $\Omega$}. The seven cases below are summarized here; each is
developed in full, with quantitative forecasts, figures, and references, in the
group's detailed document \citep{rubin2036_groupC}.

\subsection{Cosmology and extragalactic science}
\label{sec:C-cosmo}

\paragraph{High-redshift large-scale structure with medium bands}
The goal is to measure
high-redshift large-scale structure using Lyman-$\alpha$ emitters (LAEs) at
$z\gtrsim2$ selected with medium-band filters, together with LSST-selected
Lyman-break galaxies (LBGs) refined by medium-band photometry, and to perform a
$2\times2$-point analysis (LAE/LBG clustering $\times$ cosmic microwave
background, CMB, lensing) constraining
$\Omega_M$, $S_8$, neutrino mass, and $(w_0,w_a)$ with the longest available
lever arm, while accessing primordial-Universe physics ($f_{\rm NL}$) where
structure is less nonlinear. The headline figure of merit is a forecast
$\sim$4--5\% constraint on $\sigma_8$ per medium band from the LAE clustering
$\times$ Simons Observatory CMB-lensing cross-correlation, well beyond existing
measurements at this redshift. Only Rubin can detect LAEs and enable clean LBG
selection over the $\geq$11{,}000\,\degsq\ that Spec-S5
requires: \edit{the Blanco telescope is a candidate to be repurposed to host Spec-S5 itself}, the Canada-France-Hawaii
Telescope (CFHT) has a much smaller \edit{aperture than Rubin}, and Subaru/Hyper Suprime-Cam (HSC)
surveys $\sim$3$\times$ more slowly \edit{than Rubin} and is not survey-dedicated.
The concrete need is a minimal set of $\sim$7 medium bands with
$\lambda/\Delta\lambda\sim20$ spanning $2.1\lesssim z\lesssim3.5$ with continuous coverage
(LAE wavelength--redshift mapping: 400\,nm\,$\rightarrow z=2.3$; 700\,nm\,$\rightarrow z=4.8$),
at $\sim$\$1M per filter, ideally reaching $2<z<6.5$ with \emph{Roman} supplementing the
high end. Current target-selection efficiency with \edit{the Intermediate-Band Imaging Survey (IBIS) on DECam on the Blanco telescope} plus HSC
broadbands is already 75\% against DESI. The case connects tightly to the Simons
Observatory (SO; which supplies the CMB-lensing map), DESI, and the Prime Focus
Spectrograph (PFS), and, most importantly, to Spec-S5, whose LAE sample requires $\sim$4 years of Rubin
dark-time medium-band observing as a precursor: the two facilities should be
optimized jointly. The precursor HSC-\emph{Niji} survey (medium bands binned into
four ``quadrants'' MBQ1--4 to avoid slow filter exchanges; MBQ1 delivered in
2025) demonstrates the approach, and photo-$z$ forecasts indicate that adding
$\sim$16 medium bands to the six broad bands meets LSST Year 1 (Y1; and
largely Year 10, Y10) weak-lensing and large-scale-structure (LSS) photo-$z$
requirements. The chief open question is \edit{how to}
quantify the marginal gain over LSST alone.

\paragraph{Offset broad bands for photometric redshifts}
Rather than medium bands, this case adds a set of 5--7 broad bands offset in
wavelength from $ugrizy$, gaining spectral information (and thus better
per-object \photoz\ and high-$z$ target selection) while still retaining
the depth needed for transient and proper-motion science that medium bands would
sacrifice. The target is a factor of $\sim$2 reduction in photo-$z$ scatter and
outlier rate, matching the gain expected from doubling spectral resolution and
exposure time, at \$1--2M per filter. An empirical test using the Cosmic
Evolution Survey (COSMOS) 2020 catalog found that
11 offset broad bands outperform both the six-broadband and the
twelve-medium-band scenarios in photo-$z$ scatter and outlier fraction, and a
hybrid (six medium blue $+$ six broad) is nearly as good. Because it preserves
time-domain science, this is regarded as the ``workhorse'' filter option; it
connects to Spec-S5 target
selection, CMB cluster surveys (South Pole Telescope, SPT-3G$+$/SO), X-ray
clusters (\emph{Athena}), and radio weak lensing (SKA). The prior is synthetic
photometry from COSMOS-class spectral-energy-distribution (SED) catalogs
rather than new data.

\paragraph{AGN accretion-disk reverberation mapping and cosmology}
Rubin will produce the
largest homogeneous time-domain AGN sample ever assembled, turning continuum
reverberation mapping from a targeted, single-quasar technique into a
population-scale experiment across black-hole mass, accretion rate, and redshift.
Medium bands ($\lambda/\Delta\lambda\sim15$--20) placed to separate
continuum-dominated from line-dominated spectral regions, on a $\sim$1--2\,day
cadence in the Deep Drilling Fields, would measure accretion-disk sizes and
temperature profiles after removing redshift-dependent contamination, resolving
the long-standing puzzle that measured disks are larger than thin-disk theory
predicts. A contamination-controlled disk-size sample could in principle serve as
a geometric cosmological ruler complementing Type Ia supernovae. A pilot would
target continuum lags for several hundred AGN; the mature program, several
thousand. The case has natural Chilean co-leadership through Rubin, Gemini-South,
and SOAR, with the Southern Photometric Local Universe Survey (S-PLUS) and
4MOST-ChANGES as tie-ins.

\paragraph{Protocluster assembly}
A companion extragalactic case seeks a complete catalog of protoclusters at
$z<4$, tracing when the highest-mass halos assemble and how their galaxies
transform, using precise medium-band or variable-filter photometric redshifts
and intracluster light as a mass proxy, building on \emph{Euclid}, \emph{Euclid}$\times$LSST,
and Sunyaev--Zel'dovich candidate lists (LSST broadbands alone reach only
$z\lesssim1.3$--1.4).

\subsection{Local-Universe, stellar, emission-line, and exotic-filter concepts}
\label{sec:C-narrow}

\paragraph{Galactic archaeology and local dark matter}
A set of $\sim$5 medium bands ($\sim$200\,\AA\
wide; a minimal set is Ca\,\textsc{ii}
H\&K plus Mg\,$b$) sensitive to stellar metallicity and surface gravity would map
Local Group structure and formation over $\geq$70\% of the sky, identifying
streams, tidally stripped stars, and rare low-metallicity stars, and delivering a
3D distance map to the Milky Way's virial radius, a probe of small-scale dark
matter and first-star enrichment. The target is [Fe/H] to $\sim$10\% and 90\%
dwarf/giant discrimination at $g\leq25$. Offset broad bands are shown to be
insufficient here: the target features have equivalent widths of only
$\sim$5--10\,\AA, requiring true medium bands even for faint stars. The transient-cadence
cost (a 200\,\AA\ band needs $\sim$150\,s versus 30\,s to stay
background-limited) is mitigated by making exposure time a function of filter
wavelength. IBIS, \emph{Niji}, and DESI Run~2 are the precursors that will
validate the method.

\paragraph{Extended-baseline proper motions for halo dark matter}
\label{sec:C-pm}
This is a case to probe the distribution of dark matter in the Milky Way halo through
the kinematics of satellite dwarf galaxies, globular clusters, and stellar
streams, in particular by detecting and modeling the gaps in tidal streams
thought to be produced by encounters with dark-matter subhalos, thereby
constraining the subhalo mass function. Because proper-motion accuracy improves
as $t^{-3/2}$ with time baseline, continuing the survey past 2036 yields large
astrometric gains: extending proper-motion accuracy to better than
0.5\,mas\,yr$^{-1}$ at or below $r\approx24$ over the entire southern sky\edit{, within several years of continued post-LSST operation}. Rubin
will already have built a decade-long baseline over an area larger than \emph{Roman}'s,
and to depths not reached by planned astrometric surveys such as \emph{GaiaNIR},
so additional epochs \edit{compound the advantage of Rubin}. \edit{This would require no hardware change}: roughly ten to twenty
broad-band epochs per year suffice to keep the baseline growing while allowing
outlier rejection, and this case is the strongest argument for keeping some
fraction of the post-LSST cadence in \emph{broad} bands (the existing $ugrizy$
or the ``offset'' broad bands above) rather than moving entirely to medium or
narrow bands, which motivates its placement in Group~C. The prerequisite is a
demonstration, using Gaia as a prior, that LSST can measure proper motions to
the required accuracy for faint, low-proper-motion sources, plausibly by DR1 but
perhaps requiring a longer baseline. It connects to the offset-broad-band and
Galactic-archaeology cases above and to near-field cosmology programs.

\paragraph{Narrow-band H$\alpha$ emission-line imaging}
An H$\alpha$ survey \edit{that would serve} science from supernova remnants to the
star-forming filaments of large-scale structure, bridging to 4MOST spectroscopy,
exploiting Rubin's unique combination of the widest field of view and
active-optics-driven uniform image quality. It needs an H$\alpha$ filter or two
ramp filters within the existing broadband range (potentially the largest
narrow-band filter yet built), with DR1 setting the achievable redshift depth
(focus on $z<1$); a possible Chilean tie-in is 4MOST-CHANCES. \edit{Satellite
megaconstellations are a comparatively larger concern for narrow-band than for
broadband imaging, since the line signal itself is faint, so trail masking and
mitigation are more important here.}

\paragraph{Biosignature-proxy and Kuiper-belt-object triage}
Two
physics-based ``triage'' cases (explicitly \emph{not} claims of detection) ask
whether a single well-placed band sharpens a population-scale search. The first
tests whether a $\sim$710\,nm medium band (full width at half maximum, FWHM,
20--40\,nm), \edit{alternated with} $r$ and $i$, can flag a vegetation-red-edge-like
reflectance feature and separate it from haze/methane look-alikes, reducing
the number of stacked detections needed to reach $3\sigma$--$5\sigma$; the
second tests whether a $\sim$450\,nm blue-visible medium band (FWHM
$\approx$60--80\,nm) improves separation of
organic-rich, ice-rich, mixed, and weakly active Kuiper-belt-object surface
states beyond $ugrizy$ (figure of merit: $\geq$25\% gain in the pairwise
Mahalanobis distance at fixed cost)\edit{, calibrated against the New Horizons flyby measurements of the Kuiper-belt object Arrokoth as ground truth}. Both are framed as \edit{low-cost additions to the survey} that concentrate scarce
spectroscopy on the highest-information targets, and both \edit{have sensitivity} to
weather (near-simultaneous color pairs) and satellite-megaconstellation trails in
the blue.

\paragraph{Exotic filter hardware}
Three hardware concepts round out the group. The ``butcher-block'' segmented
filter array would bond 21 uniform narrow-band tiles, one
per charge-coupled-device (CCD) raft, into a single 75\,cm element, giving each raft a clean,
non-smeared bandpass \edit{that sidesteps} the fast-beam smearing problem, \edit{and} sampling
$\sim$21 redshift slices in one exposure with dithering (precedents: NASA ``butcher-block'' arrays, PAUCam). Leveraging the bandpass--angle-of-incidence
covariance, explicitly a generic technique, not a science case, uses
masks on the primary mirror to shift bandpasses by $|\Delta\lambda/\lambda|\approx0.01$
using existing glass, exploiting Rubin's telecentric coupling of pupil radius to
incidence angle (14.1$^\circ$--23.5$^\circ$). And ``ramp'' (spatially variable)
filters recur as the favored configuration for the protocluster and H$\alpha$
cases, with the caveat that their feasibility and the required new data-processing
pipeline remain to be demonstrated. A shared implication across all of these is that
current \edit{reduction} pipelines assume a discrete, unchanging filter set, so non-standard
filters would require new data-processing pipelines, a substantial but tractable
development effort.

\section{Group D: Targets of Opportunity and Multi-messenger Astronomy}
\label{sec:groupD}

\noindent\textit{Contributors:} S.~Anand, L.~Eisert,
S.~Fu, N.~Hazra, Y.~Kang, K.~Kelkar,
M.-X.~Lin, S.~MacBride, A.~Palmese,
S.~Panda, Y.~Utsumi. \\

\edit{The coming decade will bring major advances in gravitational-wave and
neutrino detection. The next-generation (XG) ground-based GW observatories (the
Einstein Telescope, ET, and Cosmic Explorer, CE), the space-based LISA, and
next-generation neutrino arrays (KM3NeT, TRIDENT, IceCube-Gen2) will eventually
increase compact-object-merger and neutrino-alert detection rates by one to two
orders of magnitude while revealing new source populations.} Rubin, with its field
of view, depth, and already-demonstrated low-latency target-of-opportunity (ToO)
infrastructure, is uniquely able to identify and characterize the electromagnetic
(EM) counterparts at these rates. \edit{Group~D} developed eight science cases,
spanning multi-messenger cosmology in the \edit{XG} gravitational-wave era,
extragalactic transient astrophysics, and Galactic multi-messenger science, all
building on Rubin's LSST-era ToO heritage and developed in full in the group's
detailed document \citep{rubin2036_groupD}.

\subsection{Bright standard sirens from ET/CE mergers}
\label{sec:D-etce}

\edit{In the nearer term, through the mid-2030s, the most likely configuration is
the Laser Interferometer Gravitational-Wave Observatory (LIGO) A$^\sharp$ upgrade \citep{ligo_asharp}: with the fifth observing run (O5)
expected to end around 2031 and A$^\sharp$ following after a few years of upgrades, a
three-detector network of the two US LIGO detectors plus LIGO-India is forecast to
yield $\sim$100 binary-neutron-star (BNS) and neutron-star--black-hole (NSBH)
mergers per year localized to $\sim$10\,\degsq\ \citep{gupta2024_networks}. Looking
further ahead, the XG observatories CE~\citep{evans2021_ce} and
ET~\citep{abac2025_et}, once operating at design sensitivity, would increase the
ground-based detection rate by one to two orders of magnitude over the current
LIGO--Virgo--KAGRA (LVK) network, to
$\sim$10$^5$--10$^6$ BNS mergers per year, of which thousands per year could be
localized to $\sim$10\,\degsq\ or less, depending on the exact network configuration
\citep{gupta2024_networks}. The goal is to extend Rubin's ToO system past LSST to
discover and localize kilonova counterparts in sufficient numbers to enable their
spectroscopic follow-up and to contribute to a percent-level $H_0$ measurement,
mapping $H(z)$ with gravitational-wave and electromagnetic data in the 2040s.} Each
bright siren requires a kilonova identified and characterized within days before it
fades; a large fraction of well-localized events fall within $\sim$1--10\,\degsq,
ideal for a single Rubin pointing. \edit{Meeting this challenge at these event rates requires Rubin's wide field of view and faint reach}
\citep{borghi2025_sirens}: only a system with Rubin's faint reach and
low-latency ToO infrastructure (demonstrated at $<5$\,min response in
commissioning; \citealt{macbride2026_too}) can triage \edit{an order of magnitude more
events than the current LVK network can}. \edit{What this science case needs} is
a rapid-cadence mode \edit{for Rubin} (minutes-to-hours) and a ToO allocation beyond the LSST-era
3\%. The prior is the LSST-era ToO performance record, with a first assessment
possible after the LVK IR1 run ($\sim$mid-2027). For context, GW170817 alone
gives $H_0$ to \edit{$\sim$14\% using the gravitational-wave detection alone~\citep{abbott2017_h0siren}, and $\sim$7\% and below in later standard-siren analyses of the same event}~\citep{hotokezaka2018_h0, palmese2024_h0siren, amsellem2026_gw170817}\edit{;}
$\sim$50 events reach $\sim$2\%~\citep{chen2018_2percent}, and a full \edit{XG}
network can \edit{measure $H_0$ to $0.2\%$ (enough to resolve the current $H_0$ tension)} within a year of
observations \citep{chen2021_multimessenger}.
In addition to kilonovae detection, standard sirens from stellar-mass black
holes detected by CE and ET could be observed using long-baseline monitoring of
the sky areas of well-localized black-hole mergers for variability exceeding
historical flares, adding a source of standard sirens that extends to larger
cosmological distances.

\subsection{Bright sirens from LISA massive-black-hole mergers}
\label{sec:D-lisa}

LISA (\edit{to be launched in} $\sim$2035) \edit{will detect} massive black-hole binary (MBHB) mergers
($\sim$10$^4$--10$^8\,\Msun$) out to $z\sim10$, with localizations from tens of
\degsq\ down to sub-\degsq, evolving over \edit{days to weeks} before merger. A
fraction of these mergers will occur in gas-rich circumbinary environments that
can produce a detectable optical
transient or quasi-periodic signature; Rubin could identify host candidates and
extend the standard-siren Hubble diagram to a much longer cosmological lever arm.
The case is for sustained, deep monitoring of LISA localization
regions \edit{over days to weeks, using} a robust photometric redshift catalog built from
LSST/\emph{Roman}/\emph{Euclid}\edit{. The expected yield is}
$\sim$1 Rubin-detectable EM counterpart per year (out to
$z\sim2$--4, model-dependent)~\citep{mangiagli2022_lisa}.

\subsection{Candidate hosts for the pulsar-timing-array background}
\label{sec:D-pta}

In the nanohertz band, the maturing pulsar-timing-array (PTA) effort
(NANOGrav, the European, Chinese, and Parkes arrays, MeerKAT, and ultimately the
SKA) is expected by the mid-2030s to resolve both the anisotropy of the
gravitational-wave background and a handful of individual nearby sources,
predicted to be bright, inspiraling supermassive black-hole binaries (SMBHBs)
superposed on a many-source background. Rubin's depth and wide field can supply a
catalog of candidate host galaxies for these nearby SMBHBs (expected to peak at
$z\sim0.1$--1, \edit{detected directly in host-galaxy images} or from associated AGN
variability) to seed the single-source templates \edit{that PTAs} use in their
searches. A confirmed association would characterize the bright foreground of the
background and constrain SMBHB evolution, and even a null result is informative;
the case requires only a sufficiently deep galaxy survey and no hardware change.

\subsection{A kilonova-demographics engine in the XG era}
\label{sec:D-kn}

Beyond individual sirens, Rubin's ToO system could build a large population of
well-sampled kilonova light curves from the ET/CE catalog, resolving how ejecta
mass, velocity, and composition depend on progenitor properties, moving beyond
the single well-characterized event (GW170817) available
today~\citep{metzger2019_kilonovae}. A population-level luminosity
function,
viewing-angle dependence, and ejecta-mass distribution can \edit{come only} from a large
photometric sample; this requires a distinct sub-day $griz$ rapid-cadence mode
over the first $\sim$1--2 weeks post-merger for faint and distant
kilonovae. Ejecta demographics feed
nuclear-astrophysics ($r$-process) modeling, and at ET/CE rates automated
light-curve classification and on-the-fly inference are required to triage which
events warrant spectroscopy.

\subsection{Lensed binary-neutron-star mergers}
\label{sec:D-lensedbns}

Building on the census of $>$50{,}000 galaxy-scale lenses that LSST will
assemble, Rubin's ToO system could search for the optical counterparts of
strongly lensed BNS mergers flagged by \edit{XG} detectors. The \edit{XG} era should raise the
\edit{BNS merger detection rate} to $\sim$100\,yr$^{-1}$, with an estimated $\sim$0.1--3 lensed events per
year; catching the rapidly-rising, intrinsically UV-bright kilonova (redshifted
into the optical, where $u$-band imaging aids discrimination from other
transients) demands slewing and imaging within the first hour, exactly the
rapid-response ToO capability this program builds. A single lensed
multi-messenger detection would be transformative, extending standard-siren
cosmology to $z\sim1$--2, well beyond the reach of unlensed events. No new
hardware is required; the case connects to \edit{several collaborations: the LVK/CE/ET, the LSST Strong Lensing, the LSST Transients and Variable Stars, and the LSST Dark Energy Science Collaboration (DESC)}.

\subsection{Neutrino electromagnetic counterparts}
\label{sec:D-nu}

By the mid-2030s, KM3NeT~\citep{km3net_loi} and TRIDENT will open the
Southern sky at TeV energies
and IceCube-Gen2~\citep{icecube_gen2} will \edit{deliver} ultra-high-energy
($>$100\,TeV) alerts,
\edit{extending uniform high-energy neutrino coverage across the whole sky, which IceCube alone cannot provide}. At conventional energies
(1--50\,TeV) neutrino error boxes are large (0.8--7\,\degsq, up to 50\,\degsq\ for
cascades), which Rubin can blanket in a few pointings to deliver \edit{potential counterparts with arcsecond localizations}, using its $z/y$-band depth to peer through dust. The needed change
is a ToO scheduler protocol that promptly triggers low-latency $z/y$ imaging on a
verified neutrino alert; a clean deep-Southern \edit{reference image and catalog of known variable sources} (a prior from
early LSST) is required to \edit{distinguish a genuine neutrino counterpart from the many unrelated variable sources in the error region}. \edit{Several aspects of the observing strategy remain to be worked out}; the case is framed as an argument that Rubin's relevance to neutrino
follow-up will grow as electromagnetic counterparts to neutrino signals are
discovered.

\subsection{Galactic supernovae}
\label{sec:D-galsn}

With the \edit{advent of next-generation neutrino observatories}, the group proposed to
leverage Rubin Observatory for rapid electromagnetic follow-up of the next
Galactic supernova, likely a core-collapse supernova (CCSN). This is a rare event
($\sim$1--2 per century) that would enable unprecedented multimessenger
characterization by combining neutrino, gravitational-wave, and optical data. The
case has two parts: first, extending Rubin's target-of-opportunity (ToO)
capabilities beyond LSST to respond to neutrino alerts (from SNEWS
\citealt{snews2020}, DUNE \citealt{dune2024_sn}, and Hyper-K
\citealt{hyperk2022}), \edit{for which} Rubin's wide field of view and depth are essential
for scanning the $\sim$2--25\,\degsq\ localization regions these detectors provide
and for catching faint or dust-obscured signals; second, building a dedicated
all-sky rapid-response mode triggered by next-generation gravitational-wave
detectors (\edit{CE and ET}, operational by the
mid-2030s), since these could detect \edit{Galactic} CCSNe anywhere on the
sky, and only a Rubin response within minutes can capture the scientifically
unique shock-breakout and pre-peak lightcurve evolution that serendipitous
discovery cannot guarantee \citep{banovetz2026_galacticsn}. Both cases require
modest additional development (leveraging existing LSST ToO infrastructure and
strategies) and near-instantaneous autonomous scheduler overrides given the
minutes-scale response window, and offer strong connections to neutrino/GW
collaborations.

\subsection{White-dwarf binaries and the SN\,Ia progenitor problem with LISA}
\label{sec:D-wdlisa}

LISA will detect a large Galactic population of double-white-dwarf (DWD) and
other ultra-compact binaries in the \edit{millihertz (mHz)} band, some with optical counterparts
(eclipsing or AM\,CVn-type accreting systems). Cross-matching LISA sky
localizations and orbital periods against Rubin's alert stream and photometric
archive, vetted with Gaia astrometry, spectroscopy, and archival
catalogs, can identify and confirm these counterparts, building on the
$\sim$50--75 systems expected to be found jointly by LISA and LSST. The signature
outcome is a merging DWD detected by LISA that is spatially and temporally
coincident with a Rubin-discovered SN\,Ia: a smoking-gun test of the
double-degenerate progenitor channel and a direct measurement of the
merger-to-explosion delay. Realizing it motivates continued rapid-ToO response
and regular deep, high-cadence monitoring of the Galactic plane after LSST.

A cross-cutting outcome of the group is the
need for direct, collaboration-level coordination of operational timelines
between Rubin and the other major facilities (gravitational-wave and neutrino
detectors, in particular): the community should avoid, for example, taking the
main camera off-sky for a multi-year upgrade during a partner experiment's
planned multi-year observing run. The group also examined the impact of weather
and climate at Cerro Pach\'on on post-LSST operations: multi-messenger \edit{ToOs}
are particularly vulnerable to weather losses, since a fading
transient cannot be re-observed, and long-term climate trends (rising integrated
water vapor and more frequent extreme El~Ni\~no events) should be folded into
post-LSST scheduling and \edit{ToO} simulations.

\section{Group E: Enabling Technology and Upgrades}
\label{sec:groupE}

\noindent\textit{Contributors:}
S.~Allen, A.~Bolton, C.~Claver, H.~Drass, A.~Drlica-Wagner, K.~M.~Fanning, M.~Hyun, J.~B.~Kalmbach, K.~Lin, S.~MacBride, A.~I.~Malz, G.~Megias Homar, A.~Nguyen, A.~A.~Plazas Malag\'on, C.~Poppett, S.~Ritz, A.~Roodman, B.~Stalder, C.~Stubbs, J.~Su\'arez-P\'erez, S.~Thomas, Y.~Utsumi \\


Group~E was concerned with the instrumental capabilities underpinning the scientific opportunities discussed by the other groups.
Proposed modifications were organized into three tiers by cost, risk, and lead time. 
The discussion focused on hardware modifications, though algorithmic advancements and on-summit edge computing were discussed briefly.
Two physical constraints were \edit{particularly} important:
(1) Rubin's fast $f/1.2$ beam  limits how narrow or spatially variable a filter can be, and 
(2) the camera is modular electronically but not mechanically, so even a single-raft swap requires the costly operation of opening the cryostat. 
Several system configuration changes were identified that could be implemented almost immediately to provide novel capabilities.
Moderate upgrades were perceived to be possible to implement on the mid-2030s timeline, and the design and fabrication of a new filter set was perceived to have the largest marginal science gain per dollar.
Major upgrades were discussed briefly, since they were not perceived to be viable on a mid-2030s timeline.

The scope of Group~E overlaps Group~C (filters) and underpins Groups~A, B, and D (image quality, fast readout, on-summit
compute). This section is best read as a summary of the technical capabilities that could enable the scientific roadmaps described in the other sections.
The full tier-by-tier discussion, together with the group's appendices summarized
in \S\ref{sec:E-actions}, is given in its detailed document
\citep{rubin2036_groupE}.

\subsection{Tier 1: configuration-only changes (\edit{minimal hardware cost,} months--1\,yr, software/configuration/calibration)}
\label{sec:E-config}

Several new capabilities require no procurement, only new operating modes,
calibration, and software, and can be interspersed with survey operations using
hardware that already exists (a spare ITL raft \edit{(a 3$\times$3 array of CCDs)} at SLAC, \edit{the LSSTComCam single-raft commissioning camera}, a spare shutter).
These include readout-mode changes (current 3.3\,s at 5--6\,e$^-$ rms; 2.2\,s
demonstrated within noise spec; row-binning to $0.2\times0.4$\,arcsec$^2$ pixels
nearly doubles readout speed); a region-of-interest (ROI) ``guider'' mode (design
$50\times50$\,pixels/amp at 9\,Hz; $\sim$80\,Mpix across the focal plane at
5\,Hz); open-shutter, continuous-readout ($\sim$10\,ms effective temporal resolution) and
stuttered readout (effective frame rates $\sim$20\,Hz); a no-filter ``light
bucket'' mode; 
and recoating the combined primary/tertiary mirror (M1M3) to improve $u$-band
throughput, where Rubin is most photon-starved.
Together these \edit{improve capabilities for} fast-transient, FRB-counterpart,
AGN-variability, Milky-Way GW-counterpart, and $\sim$10--100\,ms compact-object
light-curve science, directly serving Groups~B and~D, and the group flagged
them as the fastest, lowest-risk items on the whole list.

\subsection{Tier 2: moderate upgrades (\$10M--\$100M, $\sim$5\,yr R\&D + $\sim$5\,yr build)}
\label{sec:E-moderate}

\paragraph{Alternative filter sets.}
A new filter complement was perceived as an avenue to deliver high-impact science on a timescale that could be commensurate with the end of LSST.
The technical viability of several different alternative filter sets was discussed.
Rubin's fast $f/1.2$ beam was of central importance in these discussions. This fast beam spreads the light from a point source over a $\sim$12\,cm ($\sim$100\,cm$^2$) footprint on the filter, or conversely, a point on the filter surface is spread over $\sim$12\,cm on the focal plane. In addition, the incident angle for the fast beam ranges from $\sim$14$^\circ$--23$^\circ$, thereby limiting the minimum bandwidth of a filter. Together, these features limit both the minimum bandpass and spatial resolution.

Offset
broad bands ($\sim$5 filters, $\sim$100--150\,nm wide) are the
most straightforward, along with a single ``light bucket'' extra-broad filter; a medium-band set (dividing each broad band into 4--5
bands, IBIS/\emph{Niji}-like, $\sim$\$15--20M) \edit{would} plausibly deliver a factor-of-$\sim$2
improvement in photo-$z$ scatter plus emission-line and stellar-population
capability, all mounting into the existing filter exchanger with no focal-plane
redesign.

Narrow bands (feasible down to
$\sim$15--20\,nm, not $\sim$8\,nm) were disfavored due to low throughput but may be considered if motivated by science cases.
Notable variants were also proposed, including ``comb filters'' comprised of combinations of medium-band filters, a broad ``light bucket'' filter, and ``one filter to rule them all''
using a single, deliberately non-uniform (segmented, linear, or radial; offset, broad, medium, narrow, or comb) filter to build differential spectra of every object.
The binding constraint on both tame and exotic filter proposals is manufacturability, not optical design (\edit{i.e.,} the
fast beam and the vendor experience with the original filters), which a bounded
vendor design study, paired with quantitative photo-$z$ image simulations using
public tools such as \edit{\textsc{imSim}, SIGgi, and RAIL}, tasked to Group~C, can retire.

\paragraph{Other moderate upgrades.}
The group \edit{scoped additional upgrades in the moderate cost range:} 
a faster shutter (the current shutter opens or closes in $\sim$0.9\,s nominal, \citealt{ctn005_shutter}; a spare ``B-unit'' exists; $\sim$0.3\,s
achievable with modest safety engineering, 0.1\,s as a purpose-built stretch
goal);
a single-raft detector replacement as both standalone science (one raft
subtends $\sim$40\,arcmin, enough for a fast Galactic-plane survey) and a
pathfinder for a future focal-plane upgrade, with options spanning fast
single-layer poly-Si CCDs ($\sim$10\,fps), low-noise MAS Skipper/SiSeRO devices,
complementary-metal-oxide-semiconductor (CMOS) sensors, and near-infrared
(NIR)-sensitive Ge CCDs (to $\sim$1.4\,\micron) or InGaAs \edit{sensors};
ground-layer adaptive optics via an
actuated secondary (M2 GLAO), targeting $\sim$0.5\,arcsec delivered image quality
across the 3.5$^\circ$ field; 
a tunable-laser satellite calibrator in the spirit of \edit{the Landolt NASA mission} (part of \edit{the} PIONEER program), and an AuxTel upgrade to reach the sub-nanometer bandpass knowledge
that supernova cosmology and any new narrow/medium-band filter set will require; 
and GPU
edge computing on the summit for a $\sim$1--3\,s alert stream, real-time classification and ToO triggering (e.g., LISA follow-up), and
information-driven dynamic scheduling, described as the lowest-risk item with
the weakest dependencies on the other options.

\subsection{Tier 3: major upgrades (\$100M+, multi-year downtime)}
\label{sec:E-major}

The most ambitious options \edit{were enumerated} briefly\edit{,} but detailed discussion was deferred due to the long \edit{lead times} to deployment:
a full
focal-plane detector replacement (which necessarily begins with a pathfinder raft
that itself delivers science); 
mounting filters immediately in
front of the detectors to defeat the 12\,cm beam-footprint problem (which changes,
rather than eliminates, the filter-design problem because the beam is not
telecentric at the focal plane); 
a redesign of the camera and optics to perform a near-infrared wide-fast-deep survey;
and a spectroscopic instrument (a single raft of
integral-field units, which \edit{with} 100\,\micron\ fibers would need $\sim$1\,M
fibers and many \edit{hundreds of} spectrographs), considered only
because \edit{a dedicated Spec-S5 facility may not fully meet the Rubin community's wide-field spectroscopic needs}, and with the
strong caveat that spectroscopy would surrender Rubin's defining advantage of
wide-field, fast-slewing time-domain imaging.

\subsection{Actionable items and appendices}
\label{sec:E-actions}
Rather than a single recommendation, the group distilled its discussion into a set
of concrete, near-term actionable items with indicative owners and lead times:
bounded, fundable studies that could begin now to inform the Astro2030 process. The
configuration-only (Tier~1) items require only months and existing hardware, while
the moderate (Tier~2) items entail roughly one to three years of study before any
build decision. They span a vendor-plus-simulation filter design study and
photometric-redshift forecasts; readout-noise, ROI/guider, and
continuous/stuttered-readout development on spare hardware; a faster-shutter safety
study; a mirror-recoating throughput/UV trade study; an LSSTComCam detector-technology
pathfinder; ground-layer-turbulence characterization for GLAO; a summit GPU
edge-computing prototype; and a bandpass-calibration requirements study. The full,
itemized list with owners and lead times is given in the group's document, which
also includes two further appendices: an assessment of maintaining LSSTCam beyond
its ten-year design life (spares, obsolescence, and refurbishment), and a worked
proof-of-concept design of medium-band and H$\alpha$ filters optimized for Rubin's
fast $f/1.2$ beam. We refer the reader to \citet{rubin2036_groupE}
for the details.

\section{Chilean and Latin-American Engagement}
\label{sec:chile}

\noindent\textit{Contributors:}
S.~Anand, L.~Eisert, N.~Hazra, Y.~Kang, S.~Panda, B.~Ried Guachalla, M.~Rubio  \\

The Rubin Observatory is located \edit{on} Cerro Pachón, Chile, and Latin-American, and specifically Chilean, co-leadership was one of the two
\edit{cross-cutting considerations applied to every recommendation from each group} (\S\ref{sec:intro};
Appendix~\ref{app:prompts}), and several of \edit{the science cases developed in the studies} already
name a Chilean co-lead or tie-in (e.g., the transiting-exoplanet census and the
strong-lensing time-delay case, \S\ref{sec:groupB}; the medium-band
Galactic-archaeology and AGN cases, \S\ref{sec:groupC}). A dedicated Day-2
mini-session, ``Chilean Community and (post-)LSST Survey'', went
further and examined directly \edit{the obstacles to broader Chilean and Latin-American co-leadership}.

\edit{The consensus from the session} was that the obstacle is organizational more than
scientific. From the point of view of cosmology, for example,
Chile has comparatively few theoretical cosmologists and, as
participants noted, \edit{limited institutional participation in the DESC}, and few Chile-based
researchers currently lead Rubin-related projects. Participants argued this is
not for lack of interest, but because of disconnection between institutes, an
unresolved policy for how Rubin data rights apply to non-US, non-Chilean-national
researchers based in Chile, and the absence of \edit{a clearly publicized point of contact for who leads Rubin-related science of interest to the Chilean community}. The formal machinery that exists, an implementation
committee and the Rubin Survey Cadence Optimization Committee, SCOC, was reported to have made little visible progress since the
pandemic, and the data-center plan for Chilean community access to post-LSST
Rubin data remains undetermined.

Two existing programs were raised as models to build on: the annual La Serena Data School (hosted by NOIRLab/Gemini-South) and
Cerro Tololo Inter-American Observatory (CTIO)/Rubin site visits, both valued by participants but constrained by
scheduling that does not always match the Chilean academic calendar.
Benchmarking against other Chile-based facilities pointed to two concrete
templates: the Simons Observatory's SO-Chile initiative (research-exchange
internships of order 2--4 slots and $\sim$6 months, plus in-kind contributions,
run in partnership with Universidad de Antofagasta), and the Gemini/Chile-office
postdoctoral model (funded jointly by Gemini and a grant from Chile's National
Fund for Scientific and Technological Development, FONDECYT, or \edit{from the host}
university), which participants suggested Rubin could adapt into a joint
fellowship, and which has an international analog in bilateral postdoc programs
such as the Korea Astronomy and Space Science Institute (KASI)/KIPAC.

The session's proposed near-term next steps are: (i) \edit{encourage Rubin to publish} a clear list of points of contact and
project leadership for Rubin-related science, together with an explicit
statement of the data-rights policy for Chile-based, non-Chilean-national
researchers; (ii) reconvene the implementation and SCOC committees; (iii) pilot
a joint postdoctoral fellowship on the Gemini/Chile-office model; and (iv) hold a
Chile-based Rubin~2036 event, timed to the Chilean academic calendar rather than
to the convenience of visitors.

\section{Conclusion}
\label{sec:conclusions}

The \emph{Rubin 2036: Opportunities for Discovery} workshop gathered community participants around the question: \edit{what should be done with Rubin after the LSST?} \edit{The workshop built directly on} more than a
decade of prior community planning (Appendix~\ref{app:prior}) and \edit{was timed} to
prepare well-developed science cases for the next decadal survey (Astro2030).
Although the LSST began only in 2026, the long lead times of any facility
change, from cadence and filters to focal-plane hardware, mean that it is not too
early to start this process now. The overarching \edit{consideration} is that Rubin's
\'etendue will remain unmatched through the 2030s, so continuing to operate and
evolve the facility \edit{can be} a high-return, best-available investment rather than a
marginal extension.

The five groups reached the following main conclusions:
\begin{itemize}
\item \textbf{Group~A} (continued LSST-style surveying): a deeper, extended survey
pays off only if the faint, crowded sky can be measured, so the priority is
sustained investment in data-analysis software, computing, and machine-learning
infrastructure, such as survey-native
foundation models, rather than in new hardware (Section~\ref{sec:groupA};
\citealt{yadav2026_deepdisc}).
\item \textbf{Group~B} (alternative cadences and footprints): a flexible,
proposal-driven process (a ``metacase'') for allocating post-LSST time is the
natural organizing framework, enabling science-driven cadences from a Milky-Way
exoplanet-transit census to Type~Ia supernova cosmology, time-delay cosmography,
and second-scale transients (Section~\ref{sec:groupB}; \citealt{rubin2036_groupB}).
\item \textbf{Group~C} (narrow- and medium-band surveys and filters): new filter
sets are the highest-leverage, modest-cost evolution of the hardware, with offset
broad bands preserving time-domain science and medium and narrow bands unlocking
high-redshift large-scale structure, AGN reverberation mapping, Galactic
archaeology, and emission-line science (Section~\ref{sec:groupC};
\citealt{rubin2036_groupC}).
\item \textbf{Group~D} (targets of opportunity and multi-messenger): extending
Rubin's demonstrated target-of-opportunity system into the \edit{next-generation (XG)}
gravitational-wave and neutrino \edit{observatory} era \edit{will make} it the premier engine for multi-messenger
cosmology and astrophysics, at a modest fraction of observing time
(Section~\ref{sec:groupD}; \citealt{rubin2036_groupD}).
\item \textbf{Group~E} (enabling technology and upgrades): the instrumental options
fall into configuration-only changes available now and moderate upgrades feasible
on the mid-2030s timeline, with a new filter complement offering the largest
science return per dollar (Section~\ref{sec:groupE}; \citealt{rubin2036_groupE}).
\end{itemize}

Cross-cutting priorities recurred throughout, including opportunities for Chilean and
Latin-American co-leadership (Section~\ref{sec:chile}), the impact of weather and
climate at Cerro Pach\'on, the effect of satellite megaconstellations, and the
programmatic reality that any post-LSST program must make a compelling case to multiple funding agencies and sources within the facility's operating-cost envelope. The
science cases collected here are offered as a non-exhaustive reference point for the
community's contributions to the Astro2030 process.

Next steps include an
online follow-up seminar, to be held as a Rubin Science
Assembly\footnote{\url{https://rubinobservatory.org}} on 29 October 2026, followed
by a second in-person meeting in the series, which we aim to hold in 2027;
dedicated sessions at the LSST@ regional community meetings (e.g., LATAM, Europe, Asia) and at future Rubin
Community Workshops are also planned. The workshop's outputs, the per-group
documents and their supporting materials, are archived in the Rubin~2036 Zenodo
community\footnote{\url{https://zenodo.org/communities/rubin2036workshop/}} and linked from the workshop
website,\footnote{\url{https://rubin-observatory.github.io/Rubin-Beyond-LSST/}}
which \edit{will serve} as a persistent home for this and future meetings in the series.

\begin{acknowledgments}
This work resulted from collaborations supported by the Kavli Institute for
Particle Astrophysics and Cosmology (KIPAC) through a Workshop Program grant to
Andr\'es Alejandro Plazas Malag\'on (AAPM) for Rubin 2036 held from August 3rd to 5th,
2026, made possible by an endowment by the Kavli Foundation.

The work of AAPM was supported by the U.S. Department of Energy under contract number DE-AC02-76SF00515. AAPM thanks the Department of Physics of Harvard University and the Laboratory of Particle Astrophysics and Cosmology, the Cosmology Group at Boston University, and the Department of Physics at Washington University in St. Louis for their hospitality during the preparation of this paper.

R.~Bonito acknowledges support by the INAF Mini-Grant ``Physical properties of
Accreting young stellar objects: exploration of their light Curves and Emission
(PACE).'' A.~S.~Tramuto and R.~Bonito acknowledge support by the Rubin LSST
Discovery Alliance Science Catalyst Grant PREVIEW (PI: Bonito).

Argonne National Laboratory's work was supported by the U.S.\ Department of
Energy, Office of High Energy Physics, under contract DE-AC02-06CH11357.
\end{acknowledgments}

\appendix

\section{The Workshop Prompts}
\label{app:prompts}

Workshop participants were given prompts with questions to answer per science case to provide structure to the discussions and guide their development. 

\startlongtable
\begin{deluxetable}{l l}
\tablecaption{The common set of prompts each group used to structure and compare
its science cases.\label{tab:prompts}}
\tablehead{\colhead{Prompt} & \colhead{Question}}
\startdata
\parbox[t]{1.9in}{\raggedright Goal} & \parbox[t]{4.5in}{\raggedright In 1--2 sentences, state the specific science question and the single result that would matter. Name \emph{one} question, not a whole field.} \\[3pt]
\parbox[t]{1.9in}{\raggedright Why Rubin, why now} & \parbox[t]{4.5in}{\raggedright Say what \emph{only} Rubin can do here: its area $\times$ field-of-view $\times$ time (\aot) that no other 2036-era facility matches, and why the decision cannot wait.} \\[3pt]
\parbox[t]{1.9in}{\raggedright The skeptic's objection and our answer} & \parbox[t]{4.5in}{\raggedright Write the strongest reason a decadal panel would say \emph{no}, then a short rebuttal framed as the \emph{best available} investment, not merely a good idea.} \\[3pt]
\parbox[t]{1.9in}{\raggedright What Rubin needs} & \parbox[t]{4.5in}{\raggedright Name the concrete change (filter set / cadence / instrument / software) and a rough lead time (filters $\gtrsim$2--3\,yr; focal-plane technology $>$5\,yr).} \\[3pt]
\parbox[t]{1.9in}{\raggedright Priors} & \parbox[t]{4.5in}{\raggedright Which early-LSST result (DR3 / DR6\ldots) must be in hand before this case is airtight?} \\[3pt]
\parbox[t]{1.9in}{\raggedright Success metric / figure of merit} & \parbox[t]{4.5in}{\raggedright Give the standard metric \emph{and} a broader ``discovery-potential'' metric: what this unlocks that a traditional dark-energy figure of merit misses. Name 1--2 concrete numbers.} \\[3pt]
\parbox[t]{1.9in}{\raggedright Connections} & \parbox[t]{4.5in}{\raggedright Which other facilities, surveys, and workshop groups does this link to?} \\[3pt]
\parbox[t]{1.9in}{\raggedright Next steps} & \parbox[t]{4.5in}{\raggedright The most useful next steps after the workshop, and potential contributors.} \\[3pt]
\parbox[t]{1.9in}{\raggedright AI/ML angle} & \parbox[t]{4.5in}{\raggedright Does artificial intelligence / machine learning enable or change this case? If so, is it essential to the method or an accelerant, and what data, compute, or software would it require?} \\[3pt]
\parbox[t]{1.9in}{\raggedright Chilean / Latin-American co-leadership and engagement} & \parbox[t]{4.5in}{\raggedright Name a co-lead, data-access, or training/exchange tie-in that would advance Chilean and Latin-American participation in the case (see also \S\ref{sec:chile}).} \\
\enddata
\end{deluxetable}

The last two entries in Table~\ref{tab:prompts} (AI/ML and Chilean/Latin-American
co-leadership) are \edit{cross-cutting considerations applied in each topical area} where relevant,
alongside the eight science prompts.

Two further considerations arose from the workshop's own discussions, rather
than being specified in the original prompts, and were applied
retroactively to cases where they were relevant: the impact of weather (seeing,
cloud cover, and precipitable water vapor, including their long-term trends at
Cerro Pach\'on) on any proposed observing strategy, and the impact of satellite
megaconstellations on image quality, scheduling, and downstream data processing.

\section{Framing Talks}
\label{app:talks}

A small number of invited talks on Day~1 and Day~2 framed the discussion without
substituting for it; per the workshop's discussion-heavy design (\S\ref{sec:intro}). Slides for a subset of them are available at the workshop's website \url{https://rubin-observatory.github.io/Rubin-Beyond-LSST/pages/Rubin2036proceedings/}.

\startlongtable
\begin{deluxetable}{c l l}
\tablecaption{Invited framing talks (Day~1) and synergy talks (Day~2).\label{tab:talks}}
\tablehead{\colhead{Day} & \colhead{Speaker} & \colhead{Talk}}
\startdata
1 & A.~Plazas Malag\'on (SLAC/KIPAC/Rubin) & \parbox[t]{3.9in}{\raggedright ``Introduction'': workshop motivation, Rubin beyond the LSST, charge and goals.} \\[3pt]
1 & A.~Roodman (KIPAC/SLAC) & \parbox[t]{3.9in}{\raggedright ``Previous Work and Technical Constraints,'' surveying the studies reviewed in Appendix~\ref{app:prior}.} \\[3pt]
1 & R.~Blum (NOIRLab) & \parbox[t]{3.9in}{\raggedright ``NOIRLab's Perspective on Plans and Processes for the Future.''} \\[3pt]
1 & R.~Wechsler (KIPAC/SLAC) & \parbox[t]{3.9in}{\raggedright ``A Crystal-ball View of Astrophysics and Cosmology in 2036.''} \\[3pt]
1 & J.~Frieman (KIPAC/SLAC) & \parbox[t]{3.9in}{\raggedright ``Decadal and Funding Agency Landscape.''} \\[3pt]
2 & C.~Poppett (UC Berkeley/LBNL) & \parbox[t]{3.9in}{\raggedright ``A Next-Generation Spectroscopic Observatory,'' on the Spec-S5 concept.} \\[3pt]
2 & J.~Newman (U.~Pittsburgh) & \parbox[t]{3.9in}{\raggedright ``Spec-S5: The Concept from Snowmass.''} \\[3pt]
2 & H.~Miyatake (KMI/Nagoya) & \parbox[t]{3.9in}{\raggedright \edit{``HSC Niji'': medium-band cosmology with Subaru/HSC as a pathfinder for Rubin.}} \\[3pt]
2 & A.~Palmese (CMU) & \parbox[t]{3.9in}{\raggedright ``Gravitational-Wave Multi-messenger Astronomy in the 2030s.''} \\
\enddata
\end{deluxetable}

\section{Lightning Talks}
\label{app:lightning}

Day-1 lightning talks. The slides are available at the workshop's website \url{https://rubin-observatory.github.io/Rubin-Beyond-LSST/pages/Rubin2036proceedings/}

\startlongtable
\begin{deluxetable}{l l c}
\tabletypesize{\footnotesize}
\tablecaption{Day-1 lightning talks (2--3 minutes each), which seeded many of the science cases.\label{tab:lightning}}
\tablehead{\colhead{Speaker} & \colhead{Topic} & \colhead{Group}}
\startdata
P.~Yadav & \parbox[t]{4.2in}{\raggedright Deep-learning detection, deblending, and star/galaxy classification with DeepDISC on LSST DP1 (scene-level model matches random forests and beats morphology-only classification at faint magnitudes).} & A \\[3pt]
S.~Jha & \parbox[t]{4.2in}{\raggedright Rubin as chronicler of the time-domain sky; the case for \emph{shifted broad-band} filters (not medium/narrow) as the post-LSST workhorse, and for a proposal-based, community-driven survey model.} & B \\[3pt]
A.~Chiti & \parbox[t]{4.2in}{\raggedright Metallicity-sensitive imaging for dark-matter science (DECam MAGIC; tidal tails around Crater~II; Rubin can exceed Euclid/Roman via proper motion $+$ metallicity).} & C \\[3pt]
J.~Newman & \parbox[t]{4.2in}{\raggedright Gaining spectral information while retaining transient science with offset broad bands; COSMOS2020 test showing 11 offset bands beat both 6-broad and 12-medium scenarios in photo-$z$.} & C \\[3pt]
M.~Wood-Vasey & \parbox[t]{4.2in}{\raggedright ``A truly global Rubin'': science, community, and data questions, including regional \emph{Rubin~2036@} events and industrializing data processing.} & cross-cutting \\[3pt]
S.~MacBride & \parbox[t]{4.2in}{\raggedright GW cosmology and astrophysics in 2036: bright sirens from ET/CE binary neutron stars and LISA massive-black-hole binaries; Rubin's field of view is ideal for well-localized mergers.} & D \\[3pt]
H.~Ebina & \parbox[t]{4.2in}{\raggedright Medium-band target selection of Lyman-$\alpha$ emitters (IBIS $+$ DESI; $\sim$650 LAE\,\degsq$^{-1}$ at 75\% purity), a pathfinder for high-$z$ LSS cosmology and Spec-S5.} & C \\[3pt]
J.~F.~Su\'arez-P\'erez & \parbox[t]{4.2in}{\raggedright Unsupervised anomaly detection and transient classification (ML on 58M DESI DR2 spectra), adaptable to Rubin's $\sim$10-billion-source light-curve stream.} & cross-cutting \\[3pt]
B.~Long, N.~J.~Mason, A.~Kova\v{c}evi\'c & \parbox[t]{4.2in}{\raggedright Broadband Rubin photometry as a first-stage physical discriminator for outer-Solar-System targets (a candidate $\sim$450\,nm band).} & C \\[3pt]
A.~Kova\v{c}evi\'c, N.~J.~Mason, M.~Moore & \parbox[t]{4.2in}{\raggedright Can one 710\,nm band turn a weak red-edge signal into a decision-grade Rubin measurement? (physical triage, not biosignature confirmation).} & C \\[3pt]
S.~Birrer & \parbox[t]{4.2in}{\raggedright A strong-lensing ``crystal ball'' for Rubin~2036: an alternative cadence on the Roman $\sim$2{,}000\,\degsq\ footprint (5$\times$ cadence, 2$\times$ exposure).} & B \\[3pt]
S.~Corley & \parbox[t]{4.2in}{\raggedright The exoplanet gap: Kepler covered $\sim$0.25\% of the sky and TESS only bright stars, leaving the Galaxy's planet population unmapped.} & B \\[3pt]
A.I.~Malz & \parbox[t]{4.2in}{\raggedright ``One filter to rule them all'' plus information-theoretic cadence optimization: a spatially varying filter building differential spectra, and an improved approach to survey strategy.} & E \\[3pt]
N.~Hazra & \parbox[t]{4.2in}{\raggedright Kilonova detections leveraging Rubin's synergy with XG GW detectors: $\mathcal{O}(100)$ well-localized binary-neutron-star mergers per year, $\sim$30--100 joint ET--Rubin detections per year, informing which Einstein Telescope configuration is built.} & D \\
\enddata
\end{deluxetable}

\section{Prior Community Efforts}
\label{app:prior}

This workshop builds on more than a decade of community planning on the future of
Rubin/LSST after its main survey. We summarize that lineage here; a consistent
thread is that Rubin/LSST is a facility to be \emph{evolved, not retired}, and
that the concrete decision among ``continue as-is / modest filter change / new
instrument'' should take into consideration results from on-sky LSST data. 

\paragraph{The Cosmic Visions Dark Energy Panel (2016--2018).}
Convened by DOE High Energy Physics after the 2014 Particle Physics Project
Prioritization Panel (P5) report, the CVDE panel held
community workshops at the University of Chicago (September 2016) and LBNL
(November 2017) and produced linked reports on the science
\citep{cvde_science} and the enabling technology \citep{cvde_technology}, followed
by a Small Projects Portfolio \citep{cvde_smallprojects}. Together they framed the
completion of Stage-IV surveys (DESI, LSST) as a launching point rather than an
endpoint, arguing that great cosmological information remains at $z>1$ and on
nonlinear scales, and proposed a moderate-cost portfolio including complementary
spectroscopy and narrow-/offset-broad-band imaging to sharpen LSST photometric
redshifts. This is the intellectual origin of the ``Stage V'' framing later
adopted by Snowmass.

\paragraph{The 2019 Argonne Next-generation Instrumentation Workshop.}
The first dedicated technical workshop on what hardware follows LSSTCam
\citep{nextgen_instrumentation2019} mapped three evolution pathways: continued
imaging (possibly with new $\sim$20\,nm narrow-band filters), an alternative-sensor
focal plane (microwave kinetic inductance detectors, MKIDs; germanium CCDs;
InGaAs for near-IR reach), or conversion to a
massively multiplexed ($\sim$30{,}000-fiber) multi-object spectrograph, and
concluded, without advocacy, that new filters are the cheapest near-term option
while cryogenic IR focal planes are currently cost-prohibitive. Its
cost/performance/risk taxonomy recurs throughout the later literature and in
Group~E of this workshop.

\paragraph{The Astro2020 Decadal white papers (2019).}
Two Astro2020 submissions addressed the question directly. ``Next Generation LSST
Science'' \citep{jha2019_nextgen} laid out the continue/modify/replace scenarios,
argued LSST's imaging \'etendue will remain scientifically premier well past ten
years (especially for dynamic-sky science), and called for a mid-decade
community process. ``Future Uses of the LSST Facility'' \citep{kahn2019_futureuses},
a non-advocacy assessment by the LSST Project Science Team, quantified how coadded
depth, photometric precision, and parallax/proper-motion accuracy continue to
improve past year~10 and estimated the extended-operations cost. A companion
Astro2020 white paper made the specific science case for adding $\sim$15--20\,nm
narrow-band filters to Rubin \citep{yoachim2019_narrowband}, for stellar
astrophysics, emission-line nebulae, and improved photometric redshifts, the
direct precedent for much of Group~C.

\paragraph{Snowmass 2021.}
The dedicated Snowmass white paper ``Rubin Observatory after LSST''
\citep{blum2022_rubinafterlsst} restated the three-scenario framework and worked
through the science drivers (dark energy beyond $w_0$--$w_a$, gravity and
dark-matter tests, the $H_0$ and $\sigma_8$ tensions, strong-lensing cosmography,
GW/kilonova follow-up, dwarf-satellite proper motions, Roman/Extremely Large
Telescope (ELT) synergy),
recommending the case be revisited by year~5 of the survey. The Cosmic Frontier
topical-group reports reinforced this: the complementarity report
\citep{snowmass_complementarity} formally advocated continued Rubin operation
after LSST alongside DESI-II, CMB-S4, and a Stage-V spectroscopic facility, and
the Cosmic Frontier 04 (CF04) report \citep{snowmass_cf04}, whose \S4.5 is titled ``Opportunity: The
Rubin Observatory after LSST,'' gave the most detailed menu, escalating from
continuing as-is, to re-surveying with half-bandwidth-offset filters (roughly
halving photo-$z$ errors), to a 20--30-band low-resolution ``filter
spectrograph'' ($\sigma_z\sim0.005(1+z)$), to a full camera replacement, and
stated that it was premature to choose among them before 1--2 years of LSST data.

\paragraph{Recurring recommendations.}
Across all of these efforts the same conclusions recur: (i) Rubin should be
evolved, not retired, given its unmatched \'etendue; (ii) a standard
continue/modify-filters/replace-instrument menu; (iii) narrow-, medium-, and
offset-broad-band filters as the most consistently recommended modest-cost path;
(iv) the intrinsic value of a longer time baseline (proper motions, parallaxes,
depth) independent of any hardware change; (v) a massively multiplexed
spectrograph as the recurring ``big'' option, now largely externalized to
Spec-S5; and (vi) deferral of the final decision until real survey data are in
hand. Practical resources compiled for participants include the Rubin survey-strategy
documentation\footnote{\url{https://survey-strategy.lsst.io}} and the HSC
medium-band filter specifications\footnote{\url{https://sites.google.com/view/hsc-mb-workshop/filter-specification}}.

\section{Workshop Participants}
\label{app:participants}

Table~\ref{tab:participants} lists the workshop participants as the union of the registered participants (final registration count,
6~August 2026) with the per-group contributor rosters, case write-ups, and the
framing- and lightning-talk speakers. Names are given alphabetically by given
name, with self-reported institution/affiliation.

\startlongtable
\begin{deluxetable}{ll}
\tablecaption{Rubin 2036 Workshop Participants\label{tab:participants}}
\tablehead{\colhead{Name} & \colhead{Institution / Affiliation}}
\startdata
Aaron Roodman & SLAC National Accelerator Laboratory \\
Adam Bolton & SLAC National Lab \\
Adam Myers & University of Wyoming \\
Aditya Arunachalam & UIUC \\
Agn\`es Fert\'e & SLAC \\
Alejandro Cartes Silva & Universidad de Chile \\
Alex Broughton & KIPAC/SLAC National Lab \\
Alex Drlica-Wagner & Fermilab/UChicago \\
Alex Malz & Space Telescope Science Institute \\
Alvaro Callejas Tavera & Instituto de Astronomia UNAM \\
Anand Raichoor & APC / IN2P3 / CNRS \\
Andjelka Kovacevic & University of Belgrade Faculty of Mathematics \\
Andr\'es Plazas Malag\'on & SLAC / KIPAC / Rubin \\
Andrew Rasmussen & SLAC, KIPAC \\
Anh Nguyen & Stanford \\
Anirudh Chiti & Stanford University \\
Annika Peter & The Ohio State University \\
Anthony Englert & Brown University \\
Antonella Palmese & Carnegie Mellon University \\
Arjun Dey & NSF NOIRLab \\
Ben Sherwin & Stanford/SLAC \\
Bernardita Ried Guachalla & Stanford/SLAC \\
Brian Stalder & Rubin Observatory \\
Bryce Kalmbach & SLAC \\
Burcin Mutlu-Pakdil & Dartmouth College \\
Carlos Eduardo Falandes & INPE \\
Carrie Holt & Las Cumbres Observatory \\
Cenk Kayhan & Kayseri University \\
Charles Claver & NOIR Lab \\
Chris Stubbs & Harvard \\
Christina Williams & NOIRLab \\
Chun-Hao To & UChicago \\
Claire Poppett & UC Berkeley, Space Sciences Lab \\
Conghao Zhou & UC Santa Cruz \\
Dalya Akl & New York University, Abu Dhabi \\
David Schlegel & Berkeley Lab \\
Eric Bellm & University of Washington / Rubin Observatory \\
Francisco Castander & IEEC, ICE-CSIC, Barcelona \\
Franco Mallia & Campo Catino Observatory \\
Guillem Megias Homar & California Institute of Technology \\
Hafiz Inam Ullah & ESFM - Escuela Superior de F\'isica y Matem\'aticas - IPN \\
Haifeng Wang & LUTDLab-CWNU-China \\
Hammad Rasheed & IFIN-HH, Bucharest, Romania \\
Haruki Ebina & UC Berkeley \\
Hironao Miyatake & KMI, Nagoya University \\
Holger Drass & Rubin Operations \\
Hsin-Fang Chiang & SLAC/Rubin Observatory \\
Jay Baptista & KIPAC \\
Jeff Carlin & Rubin Observatory / NOIRLab \\
Jeffrey Newman & U. Pittsburgh \\
Jerzy Surma & Warsaw School of Economics \\
John Franklin Crenshaw & KIPAC \\
John Su\'arez-P\'erez & Tecnologico de Monterrey \\
Josh Frieman & KIPAC/SLAC \\
Justin Myles & Princeton University \\
Kenneth Lin & University of California, Berkeley / LBNL \\
Kevin Fanning & SLAC \\
Kevin Reil & SLAC/Rubin \\
Kiley Gruenberg & SLAC National Lab \\
Kouki Saito & University of Electro-Communications \\
Kshitija Kelkar & Indian Institute of Technology Bombay \\
Kyla Matan & University of Pittsburgh \\
Leonid Sajkov & Stanford/KIPAC \\
Lindsey Bleem & Argonne \\
Liyang Chen & Tsinghua University \\
Lukas Eisert & SLAC \\
Lyla Choi & Stanford \\
Maia Moore & University of Kent / University of Belgrade \\
Marcelo A B De Cicco & INMETRO \\
Masaya Yamamoto & Princeton University \\
Massimo Carpinelli & University of Milano-Bicocca \\
Maya Redden & Stanford University \\
Meng-Xiang Lin & Simon Fraser University \& CITA \\
Michael Wood-Vasey & University of Pittsburgh \\
Minhee Hyun & KIPAC / Stanford University / Rubin Observatory \\
Monica Rubio & Universidad de Chile \\
Muhammad Awais & ESFM IPN \\
Nandini Hazra & NCBJ Warsaw \\
Nicolas Mai & Paris Cit\'e University \\
Nigel J. Mason & University of Kent \\
Nitya Pandey & University of Chile and European Southern Observatory, Chile \\
Patricia Burchat & Stanford University / KIPAC \\
Paul Gontard & Stanford University \\
Paulina Venegas & NOIRLAB \\
Peter Ferguson & University of Washington \\
Phil Marshall & SLAC / Rubin \\
Phil Puxley & AURA \\
Poulomi Dey & University of Delhi \\
Prince Yadav & University of Illinois Urbana-Champaign \\
Rebecca Chen & Stanford/KIPAC/SLAC \\
Rebecca Long & University of Kent \\
Richard Delacruz & SLAC \\
Rieko Momose & NAOJ \\
Risa Wechsler & KIPAC/SLAC \\
Robert Blum & NOIRLab Rubin Observatory \\
Rosaria Bonito & INAF - Osservatorio Astronomico di Palermo, Italy \\
Ryan Foley & UC Santa Cruz \\
Ryuichi Takahashi & Hirosaki University \\
Sandrine Thomas & NOIRLab/Vera C. Rubin Observatory \\
Sarah Greenstreet & NSF NOIRLab \\
Sarang Shah & Inter-University Center for Astronomy and Astrophysics \\
Saurabh Jha & Rutgers University \\
Sean MacBride & University of Zurich \\
Seth Digel & KIPAC/SLAC \\
Shenming Fu & SLAC National Accelerator Laboratory \\
Sherman Butler-Losey & Southern Methodist University \\
Shreya Anand & Stanford \\
Shun-Sheng Li & KIPAC \\
Simon Birrer & Stony Brook University \\
Simona Mei & APC/IN2P3/UPCit\'e \\
Steve Allen & KIPAC \\
Steve Jurado & Universidad de Chile \\
Steve Margheim & NSF NOIRLab \\
Steve Ritz & UC Santa Cruz \\
Steven Dillmann & Stanford University \\
Sthabile Kolwa & University of South Africa \\
Suber Corley & Arizona State University \\
Swayamtrupta Panda & International Gemini Observatory, NSF/NOIRLab \\
Theo Schutt & Stanford/SLAC \\
Vivek Kumar Jha & NCRA-TIFR \\
Yasna Ordenes-Brice\~no & IEA, Universidad Diego Portales \\
Yijung Kang & Rubin Observatory/SLAC \\
Yousuke Utsumi & NAOJ \\
Yusuke Tampo & SAAO/UCT \\
Zeeshan Ahmed & KIPAC/SLAC \\
\enddata
\tablecomments{Self-reported at registration; affiliations may have since
changed.}
\end{deluxetable}

\section{Organizing Committees}
\label{app:committees}

A.~Roodman and A.~A.~Plazas Malag\'on served as co-chairs of the \edit{Scientific}
Organizing Committee (SOC); A.~A.~Plazas Malag\'on also chaired the Local
Organizing Committee (LOC). The SOC and LOC memberships are listed in
Tables~\ref{tab:soc} and~\ref{tab:loc}.

\startlongtable
\begin{deluxetable}{ll}
\tablecaption{\edit{Scientific} Organizing Committee (SOC)\label{tab:soc}}
\tablehead{\colhead{Name} & \colhead{Institution / Affiliation}}
\startdata
Aaron Roodman & KIPAC/SLAC \\
Alex Broughton & KIPAC \\
Alex Drlica-Wagner & Fermilab/UChicago \\
Andr\'es A.\ Plazas Malag\'on & KIPAC/SLAC/Rubin \\
Anirudh Chiti & KIPAC \\
Antonella Palmese & Carnegie Mellon University \\
Ashley Villar & CfA/Harvard \\
Bob Blum & NOIRLab \\
Chris Stubbs & Harvard \\
Chuck Claver & NOIRLab \\
Claire Poppett & LBNL \\
Elisabeth Krause & University of Arizona \\
Enrique Paillas & Universidad Diego Portales \\
Francisco Castander & ICE, Barcelona \\
Guillermo Moroni & Fermilab \\
Hironao Miyatake & Nagoya University/KMI \\
Jeff Newman & University of Pittsburgh \\
John Franklin Crenshaw & KIPAC \\
Josh Frieman & KIPAC/SLAC \\
Mario Juri\'c & University of Washington/Rubin Observatory \\
Michael Strauss & Princeton University \\
Phil Marshall & IPAC/SLAC \\
Risa Wechsler & KIPAC/SLAC \\
Steve Allen & KIPAC/SLAC \\
Tom\'as Ahumada & NOIRLab \\
Yasna Ordenes-Brice\~no & Universidad Diego Portales \\
\enddata
\end{deluxetable}

\startlongtable
\begin{deluxetable}{ll}
\tablecaption{Local Organizing Committee (LOC)\label{tab:loc}}
\tablehead{\colhead{Name} & \colhead{Institution / Affiliation}}
\startdata
Adam Bolton & SLAC \\
Alex Broughton & KIPAC \\
Andr\'es A.\ Plazas Malag\'on & SLAC/KIPAC/Rubin \\
Anirudh Chiti & KIPAC \\
Araceli Zapata & SLAC \\
Heather Shaughnessy & SLAC \\
John Franklin Crenshaw & KIPAC \\
Josh Shiode & KIPAC \\
Mar\'ia Herraez & SLAC \\
Regina Matter & SLAC \\
\enddata
\end{deluxetable}

\bibliographystyle{aasjournalv7}
\bibliography{Rubin2036_summary}

\end{document}